%% file: y3_methods.tex
\documentclass[%
 reprint,
superscriptaddress,
preprintnumbers,
nofootinbib,
 amsmath,amssymb,
 aps,
 prd,
]{revtex4-1}

\usepackage{aas_macros}
\usepackage{hyperref}
\hypersetup{colorlinks=true, citecolor=blue, urlcolor=blue, linkcolor=blue}
\usepackage{graphicx}
\usepackage{dcolumn}
\usepackage{bm}
\usepackage{xcolor}
\usepackage{enumitem}
\usepackage{setspace}
\usepackage{lineno}

\newcommand{\bea}{\begin{eqnarray}}
\newcommand{\be}{\begin{equation}}
\newcommand{\ben}{\begin{enumerate}}
\newcommand{\bi}{\begin{itemize}}
\newcommand{\eea}{\end{eqnarray}}
\newcommand{\ee}{\end{equation}}
\newcommand{\ei}{\end{itemize}}
\newcommand{\een}{\end{enumerate}}

\newcommand{\redmagic}{\texttt{redMaGiC}}
\newcommand{\maglim}{\texttt{maglim}}

\newcommand{\vell}{\boldsymbol{\ell}}

\newcommand{\om}{\Omega_\mr m}

\newcommand{\mr}{\mathrm}

\newcommand{\bs}{\boldsymbol}

\begin{document}

\preprint{DES-2020-0554}
\preprint{FERMILAB-PUB-21-240-AE}

\title[DES-Y3 3$\times$2pt model validation]{Dark Energy Survey Year 3 Results: Multi-Probe Modeling Strategy and Validation}

\input{authorlist.tex}

\date{\today}

\begin{abstract}
This paper details the modeling pipeline and validates the baseline analysis choices of the DES Year 3 joint analysis of galaxy clustering and weak lensing (a so-called ``3$\times$2pt'' analysis). These analysis choices include the specific combination of cosmological probes, priors on cosmological and systematics parameters, model parameterizations for systematic effects and related approximations, and angular scales where the model assumptions are validated. We run a large number of simulated likelihood analyses using synthetic data vectors to test the robustness of our baseline analysis. We demonstrate that the DES Year 3 modeling pipeline, including the calibrated scale cuts, is sufficiently accurate relative to the constraining power of the DES Year 3 analyses.
Our systematics mitigation strategy accounts for astrophysical systematics, such as galaxy bias, intrinsic alignments, source and lens magnification, baryonic effects, and source clustering, as well as for uncertainties in modeling the matter power spectrum, reduced shear, and estimator effects. We further demonstrate excellent agreement between two independently-developed modeling pipelines, and thus rule out any residual uncertainties due to the numerical implementation.    

\end{abstract}
\maketitle

\section{Introduction}
\label{sec:intro}
Photometric wide-field imaging surveys, such as the Dark Energy Survey (DES\footnote{www.darkenergysurvey.org/}),  Hyper Suprime Cam Subaru Strategic Project (HSC\footnote{https://hsc.mtk.nao.ac.jp/ssp/}), and the Kilo-Degree Survey (KiDS\footnote{http://www.astro-wise.org/projects/KIDS/}) have published exciting results constraining the geometry and structure growth of the Universe \citep{y1kp,HSC_cosmo,KiDS1000}. These results are leading up to data sets with ever increasing statistical precision that are due to arrive in the mid 2020s from Rubin Observatory's Legacy Survey of Space and Time (LSST\footnote{http://www.lsst.org/lsst}), the Euclid satellite\footnote{sci.esa.int/euclid/}, and the Roman Space Telescope\footnote{https://roman.gsfc.nasa.gov/}. The statistical precision of the data must be matched by the modeling accuracy on the analysis side, lest cosmological constraints incur substantial systematic bias. 

This paper is part of a series describing the methodology and results of DES Year 3 (Y3) analysis, which combines cosmic shear, galaxy-galaxy lensing and galaxy clustering measurements (together termed a 3$\times$2pt analysis) from  photometric date covering more than 4,000 square degrees, corresponding to the first 50\% of the complete DES data set. Our papers present cosmological constraints for cosmic shear \citep{y3-cosmicshear1,y3-cosmicshear2}, the combination of galaxy clustering \citep{y3-galaxyclustering} and galaxy-galaxy lensing \citep{y3-gglensing}, termed 2$\times$2pt analysis, using two different lens galaxy samples \citep{y3-2x2ptbiasmodelling, y3-2x2ptaltlensresults}, as well as the 3$\times$2pt analysis \citep{y3-3x2ptkp}. These cosmological results are enabled by extensive methodology developments at all stages of the analysis; see Appendix~A of Ref.~\citep{y3-3x2ptkp} for a summary. 

While there is extensive literature characterizing the impact of different systematic effects on future surveys in isolation, data analyses require all relevant systematic effects to be identified, modeled and mitigated in combination. The corresponding \emph{analysis choices}, which encompass probes, choice of angular scales analyzed and redshift binning of the measurements/data, choice of parameterizations, and choice of parameter priors, are highly survey- and analysis-specific \citep[e.g.,][]{y1methods,KIDS_methods}. This paper motivates and validates the theoretical modeling and analysis choices for the DES-Y3 cosmic shear, 2$\times$2pt and 3$\times$2pt analyses. We systematically define astrophysical and cosmological modeling choices, and demonstrate the robustness of our adopted baseline parameterization through simulated likelihood analyses. The resulting model parameterization, analysis choices, and inference pipeline are further validated through analyses of detailed DES mock catalogs \citep{buzzard}, which are presented in \citet{y3-simvalidation}. We note that the DES-Y3 analysis was blinded to minimize experimenter bias, and the modeling and analysis choices in this paper were finalized before unblinding. We refer to \citet{y3-3x2ptkp} for a description of the blinding protocol and post-unblinding analysis choices.

In order to identify modeling choices as comprehensively as possible, we first summarize the theoretical formalism for calculating angular 3$\times$2pt statistics, avoiding specific model choices to the extent possible, in Sect.~\ref{sec:theory}. In Sect.~\ref{sec:model} we then motivate modeling choices required to evaluate the general expressions for 3$\times$2pt statistics, which are summarized in Table~\ref{tab:model}. While each of the systematics identified in Table~\ref{tab:model} has been studied extensively in isolation before, such principled approaches to identifying relevant systematics will become especially useful with increasing model complexity. We validate the robustness of the assumed scale dependence and redshift dependence of each systematic parameterization through model stress tests in the form of simulated likelihood analyses (described in Sect.~\ref{sec:inference}). These stress tests are detailed in Sect.~\ref{sec:val} and summarized in Table~\ref{tab:model}. Our conclusions are presented in Sect.~\ref{sec:conc}.

\section{Formalism}
\label{sec:theory}
In this section we derive the theoretical formalism for computing angular 3$\times$2pt statistics, assuming general relativity, a spatially flat universe, and working to leading order in the lensing distortion. The astrophysical model choices required to evaluate these expressions will be specified in Sect.~\ref{sec:model}, and we specify the galaxy samples corresponding to the DESY3 analysis in Sect.~\ref{sec:inference}.

We use lower-case, italic-type superscripts ($i,j$) to indicate tomograpic bin indices, and lower-case Greek subscripts for vector and shear components on the flat sky with respect to the Cartesian coordinate system (e.g., $\alpha = 1,2$). Lower-case roman subscripts are used to denote specific galaxy samples ($\mathrm{s}$ for the source galaxy sample, $\mathrm{l}$ for the lens galaxy sample), while the lower-case italic subscript $g$ is used for a generic galaxy sample, $g\in(\mathrm l, \mathrm s)$.
When specifying the type of correlation function or power spectrum, we use upper-case cursive subscripts to denote a generic tracer field, e.g., $\mathcal{A,B}\in (\kappa, \delta_\mathrm{m}, \delta_{g},B, E, \mathrm{I}_E,\mathrm{I}_B$). For ease of notation, we refer to angular bins by a representative separation $\theta$, but note that all correlation functions are averaged over angular bins $\left[\theta_{\mathrm{min}},\theta_{\mathrm{max}}\right]$ (c.f. Eq.~\ref{eq:bin_average}).
\subsection{Field-Level Description}
The DES-Y3 3$\times$2pt analysis is based on the cross-correlation functions of the observed projected galaxy density contrast of the lens sample, $\delta_{\mathrm{l}}^{i}$, and shear, $\gamma_{\alpha}^{j}$, in tomographic redshift bins $i$ and $j$ of the lens and source galaxy sample respectively. We use $n_g^i(z)$ to denote the normalized tomographic redshift distribution of galaxy samples.

\subsubsection{Galaxy density}
The observed projected galaxy density contrast $\delta_{g,\mathrm{obs}}^{i}(\hat{\mathbf n})$ of galaxies from sample $g$ and tomographic bin $i$ at position $\hat{\mathbf n}$ can be written as
\begin{align}
\nonumber \delta_{g,\mathrm{obs}}^{i}(\hat{\mathbf n}) =& \underbrace{\int d\chi\, W^i_{\delta,g}\left(\chi \right)\delta_{g}^{(\rm 3D)}\left(\hat{\mathbf n} \chi, \chi\right)}_{\delta_{g,\mathrm{D}}^{i}(\hat{\mathbf n})}\\
&+ \delta_{g,\mathrm{RSD}}^{i}(\hat{\mathbf n}) + \delta_{g,\mu}^{i}(\hat{\mathbf n})\,,
\end{align}
with $\chi$ the comoving distance, and $W_{\delta,g}^i = n_g^i(z)\, d z/d\chi$ the normalized selection function of galaxies in tomographic bin $i$. Here the first term ($\delta_{g,\mathrm{D}}$) is the line-of-sight projection of the three-dimensional galaxy density contrast, $\delta_{g}^{(\rm 3D)}$, of sample $g$, and the remaining terms are the contributions from linear redshift space distortions (RSD) and magnification ($\mu$).

In the linear regime, RSD contribute to the projected galaxy density contrast through the apparent large-scale flow, with velocity $\mathbf{v}$, of galaxies across the redshift boundaries of the tomographic bins, which can be modeled as \citep{IRAS_RSD,Padmanabhan07}
\small
 \begin{align}
 \nonumber \delta_{g,\mathrm{obs}}^{i}(\hat{\mathbf n}) =&\!\int\! d\chi\, 
 W^i_{\delta,g}\left(\chi \! +\! \hat{\mathbf n}\! \cdot\!  \mathbf{v}(\hat{\mathbf n} \chi,\chi)
 \right)
 \! \left[1+\delta_{g}^{(\rm 3D)}\!\left(\hat{\mathbf n} \chi, \chi\right)\right] -1\\
 \nonumber & \;+\delta_{g,\mu}^{i}(\hat{\mathbf n})\\
\nonumber \approx&\delta_{g,\mathrm{D}}^{i}(\hat{\mathbf n})-\!  \!\int\! d\chi\, W^i_{\delta,g}\!\left(\chi \right)\frac{\partial}{\partial \chi}\! \left(\frac{\hat{\mathbf n}\! \cdot\!  \mathbf{v}(\hat{\mathbf n} \chi,\chi)}{a(\chi)H(\chi)}\! \right)+\delta_{g,\mu}^{i}(\hat{\mathbf n})\,,\\
 \label{eq:RSD}
 \end{align}
\normalsize
with $a$ the scale factor, and $H(\chi)$ the Hubble rate at redshift $z(\chi)$.

The magnification term describes the change in projected number density due to geometric dilution as well as magnification effects on galaxy flux \citep{VernerV1995, Moessner1998} and size \citep{sizebias}, which modulate the selection function. The magnitude of the latter two effects depends on the flux and size distribution of the galaxy sample, and we introduce the proportionality constant $C_g^i$ to write the magnification term as
\be
  \delta_{g,\mu}^{i}(\hat{\mathbf n}) = C_g^i \kappa_g^i(\hat{\mathbf n})\,
\label{eq:delta_mu}
\ee
where we have introduced the tomographic convergence field
 \be
 \kappa_g^i(\hat{\mathbf n})=\int d\chi \,W_{\kappa,g}^i(\chi)\delta_\mathrm{m}\left(\hat{\mathbf n} \chi, \chi\right)
 \ee
 with $\delta_\mathrm{m}$ the 3D matter density contrast, and the tomographic lens efficiency
 \be
 W_{\kappa,g}^i (\chi)= \frac{3\Omega_{\rm m} H_0^2}{2c^2}\int_\chi^{\infty} d\chi' n_g^i(\chi')
 \frac{\chi}{a(\chi)}\frac{\chi'-\chi}{\chi'}\,.
 \ee
\subsubsection{Weak Lensing}
The two components $\gamma_\alpha$ of the observed galaxy shapes are modeled as gravitational shear ($\mathrm G$) and intrinsic ellipticity. The latter is split into a spatially-coherent contribution from intrinsic galaxy alignments (IA), and stochastic shape noise $\epsilon_0$
 \be
 \gamma_{\alpha}^{j}(\hat{\mathbf n}) = \gamma_{\alpha,\mathrm G}^{j}(\hat{\mathbf n})+\gamma_{\alpha,\mathrm{IA}}^{j}(\hat{\mathbf n})+\epsilon_{\alpha,0}^j(\hat{\mathbf n})\,.
 \ee
As shape noise contributes to the covariance but not to the mean two-point correlation function signal, we do not include it in the mean model predictions described here and defer to \citet{y3-cosmicshear1,y3-covariances} for details on the covariance modeling.

 The gravitational shear can be calculated as
 \be
 \gamma_{1,\mathrm G} = \left(\Psi_{11}-\Psi_{22}\right)/2\,,\;\gamma_{2,\mathrm G} = \left(\Psi_{12}+\Psi_{21}\right)/2\,,
 \label{eq:shear_components}
 \ee
 with $\Psi$ the lensing distortion tensor, which to leading order in the lensing deflection can be calculated as 
 \be
 \Psi_{\alpha\beta}^i(\hat{\mathbf n}) = 2 \int d \chi W_{\kappa,\mathrm{s}}^i(\chi)\,\chi\,\Phi_{,\alpha\beta}(\hat{\mathbf n} \chi,\chi)\,.
 \label{eq:Psi}
 \ee
 Here $\Phi_{,\alpha}$ are spatial transverse derivatives of the 3D potential $\Phi$.

 Again to leading order in the lensing deflection, in Fourier space the gravitational shear field with respect to the Cartesian coordinate system can be related to the convergence $\kappa_s^i$ by
 \begin{equation}
 \gamma^{i}_{\alpha,\mathrm G}(\vell) =T_\alpha(\vell)\kappa_{\mathrm s}^{i}(\vell)\,,
 \end{equation}
 with $\mathbf{T}(\vell) \equiv (\cos(2\phi_\ell), \sin(2\phi_\ell))$, where $\phi_\ell$ is the angle of the $\vell$ vector from the $\ell_x$ axis. 

 The intrinsic alignment contribution to the observed galaxy shear field is a projection of the 3D field $\tilde{\gamma}_\mathrm{IA}$ 
 weighted by the source galaxy redshift distribution
\be
 \gamma_{\alpha,\mathrm{IA}}^i(\hat{\mathbf n}) = \int d \chi \, W_{\delta,\rm{s}}^i(\chi) \tilde{\gamma}_{\alpha,\mathrm{IA}}\left(\hat{\mathbf n} \chi, \chi\right)\,.
\ee
The 3D intrinsic alignment field, $\tilde{\gamma}_{\mathrm{IA}}$, is specified by the choice of intrinsic alignment model, c.f. Sect.~\ref{sec:choices}.
 
The shear field, including the intrinsic alignment contribution, is decomposed into E- and B-modes as 
 \begin{equation}
 \gamma^i_{E/B}(\bm{\ell}) = \left(\delta_{\alpha\beta},\epsilon_{\alpha\beta}\right)T_\alpha(\bm{\ell})\gamma^i_\beta(\bm{\ell})\,, 
 \label{eq:EB}
 \end{equation}
 with $\epsilon_{\alpha\beta}$ the two-dimensional Levi-Civita tensor.
 To leading order, the shear E-mode is given by
 \begin{align}
\nonumber \gamma^i_{E}(\bm{\ell}) = &
 T_\alpha(\bm{\ell})\left(T_\alpha(\bm{\ell})\kappa_\mathrm{s}^{i}(\bm{\ell})
 +\gamma^i_{\alpha,\mathrm{IA}}(\bm{\ell})\right)\\
  = &\kappa_\mathrm{s}^{i}(\bm{\ell})
  +T_\alpha(\bm{\ell})\gamma^i_{\alpha,\mathrm{IA}}(\bm{\ell})\,.
 \end{align}
 To leading order, weak lensing does not produce B-modes and the B-mode shear $\gamma_\mathrm{B}$ is given by
 \begin{equation}
 \gamma^i_{B}(\bm{\ell}) = \epsilon_{\alpha\beta}T_\alpha(\bm{\ell})\gamma^i_{\beta,\mathrm{IA}}(\bm{\ell})\,.
 \end{equation}
 In the following, we set $\bm \ell = (\ell,0)$ without loss of generality.
 \subsection{Angular 2-pt Statistics}
\subsubsection{Angular Power Spectra}
In order to evaluate the angular (cross-) power spectra of the observed density contrast and shear E/B-modes, we first expand the observed fields into their physical components. Then the shear power spectra can be written as
 \begin{align}
\nonumber C^{ij}_{EE}(\ell) = & C^{ij}_{\kappa_{\rm{s}} \kappa_{\rm{s}}}(\ell) + C^{ij}_{\kappa_{\rm{s}} \mathrm{I}_E}(\ell) + C^{ji}_{\kappa_{\rm{s}} \mathrm{I}_E}(\ell)+C^{ij}_{\mathrm{I}_{E} \mathrm{I}_{E}}(\ell)\,, 
 \\
 C^{ij}_{{BB}}(\ell) = & C^{ij}_{\mathrm{I}_B\mathrm{I}_B}(\ell)\,,
 \end{align}
 where we used $\mathrm{I}_{E/B}$ as short-hand notation for the E/B-mode decomposition of intrinsic alignments, ${\gamma}_{E/B,\mathrm{IA}}$.
 
 Similarly, we model the observed galaxy-galaxy lensing power spectrum as
\be
 C^{ij}_{\delta_{g,\mathrm{obs}}\mathrm{E}}(\ell) = C^{ij}_{\delta_{g,\mathrm{D}}\kappa_{\rm s}}(\ell) + C^{ij}_{\delta_{g,\mathrm{D}}\mathrm{I_E}}(\ell) +  C^{ij}_{\delta_{g,\mu}\kappa}(\ell) + C^{ij}_{\delta_{g,\mu}\mathrm{I_E}}(\ell)\,,
\ee
where we omitted the RSD term, which is negligible for the DES-Y3 tomographic lens bin choices \citep{Fang_nonlimber}.

With the exception of the galaxy clustering power spectrum ($C^{ii}_{\delta_{g,\mathrm{obs}}\delta_{g,\mathrm{obs}}} (\ell)$, c.f. Sect.~\ref{sec:Cgg}), we calculate the angular cross-power spectrum between two fields $\mathcal{A,B}$ using the Limber approximation  \begin{align}
     C_{\mathcal{AB}}^{ij}(\ell) = \int d\chi \frac{W_A^i(\chi)W_B^j(\chi)}{\chi^2}P_{\mathcal{AB}}\left(k = \frac{\ell+0.5}{\chi},z(\chi)\right)\,,
 \end{align}
with $P_{\mathcal{AB}}$ the corresponding three-dimensional power spectrum, which is specified by the model choices detailed in Sect.~\ref{sec:model}. Ref.~\cite{Fang_nonlimber} demonstrate that using the Limber approximation for galaxy clustering may cause significant systematic biases in the inferred parameters, and show that this approximation is sufficient for galaxy--galaxy lensing and cosmic shear beyond the accuracy of the DES-Y3 analysis. The expressions for the galaxy clustering power spectrum without the Limber approximation become more tractable with the model parameterization choices in Sect.~\ref{sec:choices} and we summarize their evaluation in Sect.~\ref{sec:Cgg}.

\subsubsection{Angular 2-pt Correlation Functions}
The angular two-point correlation functions for galaxy clustering $w^i(\theta)$, galaxy--galaxy lensing $\gamma_t^{ij}(\theta)$, and cosmic shear $\xi_{+/-}^{ij}(\theta)$, are related to the angular power spectra via the transformations
 \begin{align}
     w^i(\theta) =& \sum_\ell \frac{2\ell+1}{4\pi}P_\ell(\cos\theta) C^{ii}_{\delta_{\mathrm{l,obs}}\delta_{\mathrm{l,obs}}}(\ell)~,\label{eq:transform_w}\\
     \gamma_t^{ij}(\theta) =& \sum_\ell \frac{2\ell+1}{4\pi\ell(\ell+1)}P^2_\ell(\cos\theta) C^{ij}_{\delta_{\mathrm{l,obs}}\mathrm{E}}(\ell)~,\\
     \nonumber \xi_{\pm}^{ij}(\theta) =& \sum_\ell\frac{2\ell+1}{2\pi\ell^2(\ell+1)^2}[G_{\ell,2}^+(\cos\theta)\pm G_{\ell,2}^-(\cos\theta)]\\
     &\times \left[ C^{ij}_{EE}(\ell)\pm C^{ij}_{BB}(\ell)\right]~,\label{eq:transform_xi}
 \end{align}where $P_\ell$ and $P_\ell^2$ are the Legendre polynomials and the associated Legendre polynomials, $G_{\ell,m}^{+/-}$ are given by Eq.~(4.19) of \cite{1996astro.ph..9149S}.
We calculate the correlation functions within an angular bin $[\theta_{\rm min},\theta_{\rm max}]$, by carrying out the average over the angular bin, i.e., replacing $P_\ell(\cos\theta)$, $P_\ell^2(\cos\theta)$ and $[G_{\ell,2}^+(\cos\theta)\pm G_{\ell,2}^-(\cos\theta)]$ with their bin-averaged versions $\overline{P_\ell}$, $\overline{P^2_\ell}$ and $\overline{G_{\ell,2}^+\pm G_{\ell,2}^-}$, defined by
 \begin{align}
     &\overline{P_\ell}\left(\theta_{\rm min},\theta_{\rm max}\right) \equiv \frac{\int_{\cos\theta_{\rm min}}^{\cos\theta_{\rm max}}dx\,P_\ell(x)}{\cos\theta_{\rm max}-\cos\theta_{\rm min}}\nonumber\\
     &\;\;\;\;\;\;\;\;\;\;\;\;\;\;\;\;\;\;\;\;\;\;\;\;\;\;=\frac{[P_{\ell+1}(x)-P_{\ell-1}(x)]_{\cos\theta_{\rm min}}^{\cos\theta_{\rm max}}}{(2\ell+1)(\cos\theta_{\rm max}-\cos\theta_{\rm min})}~,
\label{eq:bin_average}
 \end{align}
and analogously for $\overline{P^2_\ell}$ and $\overline{G_{\ell,2}^+\pm G_{\ell,2}^-}$,
for which analytic expressions can be found in Ref.~\citep{y3-covariances}.

\begin{table*}
\begin{tabular}{l | l | l | l | l }
Model Ingredient & Baseline Choice & Test of $k$-dependence& Test of $z$-dependence& Validation\\
\hline\hline
$P_{\rm{mm}}$  & gravity-only  & baryons+AGN feedback & baryons+AGN feedback& Sect.~\ref{sec:scale_cuts}\\
\hline
$P_{\rm{mm}}$ & halofit fitting function & higher accuracy emulators & higher accuracy emulators & Fig.~\ref{fig:test_mm}\\
\hline
galaxy bias  & linear bias $b_1^i$ & perturbative bias & passive evolution of $b_1$& Sect.~\ref{sec:scale_cuts},\ref{sec:stress_tests}\\
&per tomographic bin, Eq.~\ref{eq:PgX} & & &\\
\hline
IA & TATT (Eq.~\ref{eq:TATT}) & - & (extrapolated observations) &Sect.~\ref{sec:stress_tests}\\
&  with power-law $z$-evolution &&&\\
\hline
lensing modeling & first-order in distortion & next-to-leading order & - &  Fig.~\ref{fig:test_nlo}\\
\hline
\end{tabular}
\caption{Summary of model choices and their validation tests presented in this analysis.}
\label{tab:model}
\end{table*}
\section{Baseline Model Parameterization}
 \label{sec:model}
The formalism for calculating angular 2pt-statistics described in Sect.~\ref{sec:theory} makes few model assumptions besides a spatially-flat cosmology and working to leading order in the lensing deflection. In order to evaluate these expressions for angular two-point statistics, we now specify model prescriptions for the 3D power spectra $P_{AB}$ that include nonlinear structure formation and astrophysics. This section describes our model \emph{choices} for the DES-Y3 baseline analyses in the domain of flat $\nu \mathrm{wCDM}$ cosmologies.
 
Calculations of background evolution and transfer functions use the Boltzmann codes \texttt{CAMB} \citep{Lewis:1999bs,Lewis:2002ah} and \texttt{CLASS} \citep{CLASSII}, which are in excellent agreement at the level of accuracy of this analysis as demonstrated in Sect.~\ref{sec:code}. 

For the model validation presented here, we do not include the modeling and marginalization of observational systematics, which is a conservative choice given that it requires a more stringent performance of the theoretical model.
 \subsection{Theoretical Modeling Choices}
 \label{sec:choices}
 \paragraph*{Matter power spectrum} 
 To model the nonlinear matter power spectrum $P_{\mathrm{mm}}$, we adopt the \citet{Takahashi12} recalibration of the \textsc{halofit} fitting formula \citep{halofit} for the gravity-only matter power spectrum, including the \citet{Bird_halofit} prescription for the impact of massive neutrinos. This model has two well-known deficiencies:
 \begin{itemize}
 \item The matter power spectrum model does not account for non-gravitational forces, such as the impact of baryons. 
 \item As any fitting formula, this model has finite accuracy in representing the true gravity-only power spectrum.
 \end{itemize} 
We mitigate the model incompleteness through scale cuts that exclude the impact of strong baryonic feedback models, as detailed in Sect.~\ref{sec:scale_cuts}. In order to quantify the model accuracy we compare \textsc{halofit} against more recent matter power spectrum emulators, which are based on larger and higher resolution simulations and thus more accurate (c.f. Sect.~\ref{sec:stress_tests}).
\paragraph*{Galaxy Bias}
For the baseline analysis, we adopt a linear bias ($b_1$) prescription relative to the nonlinear matter density, such that the cross power spectrum between galaxy density and field $\mathcal{A}$ is given by
 \be
 P_{\delta_{g} \mathcal{A}}(k,z) = b_{1,g}(z) P_{\mathrm{m} \mathcal{A}}(k,z)\,.
 \label{eq:PgX}
 \ee
Here we ignored stochastic bias contributions to the galaxy power spectrum $P_{\delta_{g}\delta_{g}}$, which at leading order affect only to the zero-lag correlation function.
Furthermore, we model the redshift dependence with one free parameter $b_{1,g}^i$ per tomographic bin per galaxy sample $g$, neglecting evolution within tomographic bins.

This model choice neglects the known scale dependence of galaxy clustering due to higher-order biasing, which we mitigate through scale cuts (c.f. Sect.~\ref{sec:scale_cuts}). We further show in Sect.~\ref{sec:stress_tests} that neither neglecting the redshift evolution of linear bias within tomographic bins nor neglecting scale dependence of galaxy bias due to massive neutrinos biases the DES-Y3 analyses. These bias modeling assumptions are further validated by applying the baseline analysis model to mock catalogs \citep{y3-simvalidation}.
 \paragraph*{Intrinsic Alignments}
 As motivated in detail in Ref.~\citep{y3-cosmicshear2}, we adopt the `tidal alignment and tidal torquing' (TATT) model \citep{TATT} as the baseline intrinsic alignment model for the DES-Y3 analyses. Here, we provide a brief summary of the model. We write the intrinsic galaxy shape field, measured at the location of source galaxies, as an expansion in the density and tidal tensor $s_{ab}$, which can be decomposed into components $s_{\alpha}$ as with the cosmic shear field:
\begin{align}
\tilde{\gamma}_{\alpha,\mathrm{IA}} = A_1 s_{\alpha} +A_{1\delta} \delta_\mathrm{m} s_{\alpha} + A_2 \left(s\times s\right)_{\alpha} + \cdots\,.
\label{eq:TATT}
\end{align}
In this expansion, the first linear term, with $A_1$, corresponds to the well-studied ``nonlinear linear alignment'' model \citep[NLA,][]{Catelan2001,Hirata_IA,Bridle_NLA}. The second term captures the impact of source density weighting
\citep{Blazek15}, and together the two comprise the `tidal alignment' component. The third term, quadratic in the tidal field, captures the impact of tidal torquing \citep{Catelan2001,lee2001}.
 
Based on these alignment processes, TATT prescribes the scale dependence, and parts of the redshift evolution, of the intrinsic alignment E/B-mode power spectra $P_{\mathrm{I}_{E} \mathrm{I}_{E}},P_{\mathrm{I}_{B} \mathrm{I}_{B}}$ and the cross-power spectrum between matter density and intrinsic alignment E-mode $P_{\mathrm{m I}_E}$. Expressions for these power spectra are given in Ref.~\citep{TATT}.

At fixed redshift, the TATT power spectra depend on three amplitude parameters, NLA amplitude $A_1$, tidal torquing amplitude $A_2$, and an amplitude for the density weighting term, captured by an effective source bias $b_\mathrm{ta}$, such that $A_{1\delta} = b_\mathrm{ta} A_1$ in Eq.~\ref{eq:TATT}.
Note that in the limit $A_2, b_\mathrm{ta} \rightarrow 0$, TATT reduces to NLA. All three parameters depend on the source sample selection, and may depend on redshift. For the DES-Y3 baseline model we adopt a five parameter model, assuming $b_\mathrm{ta}$ to be constant in redshift and choosing to parameterize the redshift evolution of $A_1$ and $A_2$ as power laws with exponents $\eta_{1,2}$. In detail, the prefactors are given by
\begin{align}
A_1(z) =& -a_1\bar{C}_1\frac{\rho_\mathrm{crit}\Omega_{\rm m}}{D(z)}\left(\frac{1+z}{1+z_0}\right)^{\eta_1}\\
A_2(z) =& 5a_2\bar{C}_1\frac{\rho_\mathrm{crit}\Omega_{\rm m}}{D(z)^2}\left(\frac{1+z}{1+z_0}\right)^{\eta_2}\,
\end{align}
with pivot redshift $z_0$ corresponding to the mean redshift of the source sample, $\bar{C}_1$ a normalization constant, which by convention is fixed to $\bar{C}_1 = 5\times10^{-14} M_\odot h^{-2}\mathrm{Mpc}^2$, and $D(z)$ the linear growth factor. This power law redshift dependence is a common parameterization in recent analyses employing only the NLA-IA model as well \citep[][though the latter do not include redshift evolution of $A_1$ in their baseline analysis]{Troxel18,Hikage19,HSC_cosmo,Asgari_1000}.
We discuss the limitations of these parameterization choices in Sect.~\ref{sec:stress_tests}.
The evaluation of  perturbation theory kernels required for the TATT model, as well as similar calculations for nonlinear biasing (Sect.~\ref{sec:scale_cuts}), were performed using the {\tt FAST-PT} algorithm \citep{mcewen16,fang17}.

\paragraph*{Magnification}
We model the lensing bias coefficient $C_g^i$ introduced in Eq.~\ref{eq:delta_mu} to include the geometric dilution and the modulation of galaxy flux and size selection \citep{VernerV1995, Moessner1998,sizebias}, 
\be
C_g^i = 5 \frac{\partial \ln n_g^i}{\partial m}\bigg\rvert_{m_\mathrm{lim},r_\mathrm{lim}} +\frac{\partial \ln n_g^i}{\partial \ln r}\bigg\rvert_{m_\mathrm{lim},r_\mathrm{lim}} -2,
\ee
where the logarithmic derivatives are the slope of the luminosity and size distribution at the sample selection limit.
The values of these lensing bias coefficients are estimated from the data \citep{y3-2x2ptmagnification}, and are held fixed in the cosmology analysis. The robustness of this assumption, as well as potential redshift evolution effects within tomographic bins, are discussed in Sect.~\ref{sec:stress_tests} and further validated on mock catalogs \citep{y3-simvalidation}.
 \paragraph*{Nonlocal shear}
As shear is a nonlocal quantity, the shear two-point (cross-)correlation functions at separation $\theta$ also depend on the density distribution at scales smaller than $\theta$. For cosmic shear, this is included in validation of the nonlinear matter power spectrum modeling in Sect.~\ref{sec:stress_tests}. For galaxy-galaxy lensing, the one-halo term contribution to the tangential shear signal is non-negligible at scales far beyond the projected halo size. This effect can be mitigated by transforming the tangential shear into statistics that remove small-scale information \citep{Baldauf_2010,Mandelbaum_2013,park2020localizing}, through scale-cuts \citep{y1methods,y1kp}, or by including it in the model.
In this analysis, we adopt the point-mass marginalization scheme of \citet{MacCrann_2019}, which analytically marginalizes the tangential shear contribution of an enclosed mass and only requires priors on the enclosed mass, but no additional fit parameters.

Adapting the original expressions \citep{MacCrann_2019} to our notation, the tangential shear signal of an excess mass $B$, enclosed within the transverse scale corresponding to the smallest scale at which the correlation function is measured, is given by
 \be
 \Delta \gamma_\mathrm{t}^{ij}(\theta) = \frac{1}{\rho_\mathrm{crit} \om} \int d\chi W_{\delta_\mathrm{l}}^i(\chi) W_\kappa^j(\chi) \frac{B^i}{d_\mathrm{A}(\chi)^2 \theta^2} \equiv \frac{B^i \beta^{ij}}{\theta^2}
 \label{eq:pm}
 \ee
with $\rho_\mathrm{crit}$ the critical density, and $d_\mathrm{A}(\chi)$ the angular-diameter distance, and where we have neglected the potential redshift evolution of $B$ within the narrow tomographic lens bins of the DES-Y3 analysis\footnote{But see Ref.~\citep{y3-2x2ptbiasmodelling} for a test of this assumption.}.
Analytic marginalization over $B^i$ with a Gaussian prior $\sigma(B^i)$ modifies the data covariance $\mathbf C$ as
 \be
\mathbf C \rightarrow \mathbf C +\sum_{i=1}^{N_{z,\rm{l}}} \sigma^2\left(B^i\right)\,\mathbf{t}^i\otimes \mathbf{t}^i\,.
 \ee
 with $N_{z,\rm{l}}$ the number of tomographic lens bins, and $\mathbf{t}^i$ a vector with length corresponding to the number of data points and elements
\begin{linenomath*}
\begin{equation}
    \bigg(\mathbf{t}^{i} \bigg)_{a} = \begin{cases}
0 & \parbox{5cm}{if $a$-th element is not $\gamma_{\rm{t}}$, or if lens-redshift of $a$-th element $\neq i$} \\  
\\
\beta^{ij}\theta_{a}^{-2} &\text{otherwise}
\end{cases}
\end{equation}
\end{linenomath*}
where the expression for $\beta^{ij}$ is given in Eq.~\ref{eq:pm}.
As detailed in Ref.~\citep{y3-2x2ptbiasmodelling}, the nonlinear galaxy--matter correlation function in the one- to two-halo transition regime contributes substantially to the enclosed excess mass, and the prior on $B^i$ cannot be informed by the typical mass scale of the host halos. We adopt $\sigma(B) = 10^{17} M_\odot/h$. 
 
 \subsection{Evaluation of non-Limber Integrals}
\label{sec:Cgg}
Within the baseline model parameterization specified above, and restricting to $\nu w$CDM cosmologies, we can now express the computation of galaxy clustering power spectra in relatively compact form.

We follow \citet{Fang_nonlimber} for the non-Limber computation of galaxy clustering power spectra. Adapting their notation to the conventions in this paper \citep[also see][]{Chisari_CCL}, we define
 \begin{align}
 &\Delta^i_{g,\rm{D}}(k,\ell) = \int d\chi\,W^i_{\delta,g}(\chi) T_{\delta_{g}}(k,z(\chi)) j_\ell(k\chi)~,\\
 &\Delta_{g,\rm{RSD}}^i(k,\ell) = -\int d\chi\,f(z(\chi)) W_{\delta,g}^i(\chi)T_\delta(k,z(\chi))j_\ell''(k\chi)~,\\
 &\Delta_{g,\mu}^i (k,\ell) = \frac{\ell(\ell+1)}{k^2}C^i_{g} \int\frac{d\chi}{\chi^2} W_{\kappa,g}(\chi)T_\delta(k,z(\chi))j_\ell(k\chi)~,
 \end{align}
 where $f(z)$ is the logarithmic growth rate, $T_\delta(k,z)$ is the matter density perturbation transfer function, and $T_{\delta_{g}}$ is the transfer function of galaxy density perturbations, which assuming linear galaxy bias is given by $T_{\delta_{g}}(k,z)\approx b_{1,g}(z) T_\delta(k,z)$.

 The angular galaxy clustering power spectrum can then be written as 
 \begin{equation}
 C^{ij}_{\delta_{g,\rm{obs}}\delta_{g,\rm{obs}}}(\ell) = \frac{2}{\pi}\int\frac{dk}{k}k^3P_\Phi(k)\Delta^i_{g,\rm{obs}}(k,\ell)\Delta^j_{g,\rm{obs}}(k,\ell)~,
 \label{eq:Cl-gg}
 \end{equation}
 where $\Delta^i_{g,\rm{obs}}(k,\ell)=\Delta^i_{g,\rm D}(k,\ell)+\Delta^i_{g,\rm{RSD}}(k,\ell)+\Delta^i_{g,\mu}(k,\ell)$, and $P_\Phi(k)$ is the primordial matter power spectrum.

The expansion of the product of the different $\Delta_A(\ell)$'s leads to integrals containing two Bessel functions and their derivatives. For example, the ``DD'' term is
 \begin{align}
  \nonumber   C_{\delta_{g,\rm D}\delta_{g,\rm D}}^{ij} (\ell)=&\frac{2}{\pi}\int d \chi_1\,W^i_{\delta,{g}}(\chi_1)\int d\chi_2\,W^j_{\delta,g}(\chi_2)\\
     &\int\frac{dk}{k}k^3 P_{gg}(k,\chi_1,\chi_2)j_\ell(k\chi_1)j_\ell(k\chi_2)~,
 \label{eq:Cl-DD}
 \end{align}
 where $P_{gg}(k,\chi_1,\chi_2)$ is the unequal time, nonlinear galaxy power spectrum , while components involving the RSD have integrands containing $j_\ell j_\ell''$ or $j_\ell'' j_\ell''~$. In order to evaluate the unequal time expressions, we separate the linear part $b_{1,g}^2 P_{\rm lin}(k,\chi_1,\chi_2)$ and nonlinear contribution $[P_{gg}-b_{1,g}^2 P_{\rm lin}](k,\chi_1,\chi_2)$ of the power spectrum.
As the nonlinear contribution is significant only on small scales where the Limber approximation is sufficiently accurate, we can rewrite Eq.~(\ref{eq:Cl-DD}) as
\begin{widetext}
\begin{align}
 \nonumber   C_{\delta_{g,\rm D}\delta_{g,\rm D}}^{ij} (\ell)
    =& \int d\chi\, \frac{ W_{\delta,g}^i(\chi)W_{\delta,g}^j(\chi)}{\chi^2}
   \left[P_{gg}\left(\frac{\ell+0.5}{\chi},\chi\right)-b_{1,g}^i b_{1,g}^jP_{\rm lin}\left(\frac{\ell+0.5}{\chi},\chi\right)\right]\\
    +&\frac{2}{\pi}\int d\chi_1\,b_{1,g}^i W_{\delta,g}^i(\chi_1) D(z_1))\int d\chi_2\,b_{1,g}^j W_{\delta,g}^j(\chi_2)D(z_2) \int\frac{dk}{k}k^3 P_{\rm lin}(k,0)j_\ell(k\chi_1)j_\ell(k\chi_2)\,,
\label{eq:Cl-DD_rewrite}
\end{align}
\end{widetext}
where we have factorized the time dependence of the linear power spectrum.  We use the generalized FFTLog algorithm\footnote{\url{https://github.com/xfangcosmo/FFTLog-and-beyond}} developed in \cite{Fang_nonlimber}  to evaluate the full expression for Eq.~(\ref{eq:Cl-gg}) including RSD and magnification terms.

 \section{Likelihood Analysis Setup}
 \label{sec:inference}
The numerical implementation and scientific accuracy of our baseline model are validated through simulated likelihood analyses. We briefly summarize the likelihood analysis methodology in this section. 

\subsection{Inference and Pipeline Validation}
\label{sec:code}
We set up simulated likelihood analyses using synthetic `data' vector ${\mathbf D} \equiv\{{w}^i(\theta),{\gamma}_\mathrm{t}^{ij}(\theta),{\xi}_\pm^{ij}(\theta)\}$, generated at a fiducial cosmology, and the theoretical baseline model prediction as a function of model parameters $\mathbf p$, $\mathbf M(\mathbf p) \equiv \{w^i(\theta,\mathbf p),\gamma_\mathrm{t}^{ij}(\theta,\mathbf p),\xi_\pm^{ij}(\theta,\mathbf p)\}$ assuming a Gaussian likelihood \citep[see e.g.][for tests of the validity of this assumption in the context of cosmic shear]{lhe20}
\be
\ln \mathcal{L}(\mathbf D | \mathbf p) \propto -\frac{1}{2}\left[\left({\mathbf D}-\mathbf M(\mathbf p)\right)^\tau \mathbf C^{-1}\left({\mathbf D}-\mathbf M(\mathbf p)\right)\right]\,,
\ee
where the covariance is computed using the halo model in \textsc{CosmoLike} \citep{kre17,CosmoCov} and we refer to \cite{y3-covariances} for a detailed description of the covariance validation. For clarity, we omit the parameter argument of the model two-point functions in the following. The parameter priors assumed in the likelihood analysis and the fiducial parameter values, used to compute the (synthetic) data vector $\mathbf D$,  are summarized in Table \ref{tab:params_all}. 

The `data' is computed using preliminary DES-Y3-like redshift distributions. The source sample consists of four tomographic bins with broad redshift support extending to $z\sim1.5$ \citep{y3-cosmicshear1,y3-sompz}. The DES-Y3 considers two different lens samples
\begin{itemize}
\item \redmagic\ \citep{Rozo16}, a luminosity-threshold sample of red sequence galaxies with constant comoving density, consisting of five tomographic lens bins with redshift boundaries $[0.15,0.35]$, $[0.35,0.5]$, $[0.5,0.65]$, $[0.65,0.8]$, $[0.8,0.9]$ \citep{y3-galaxyclustering}.
\item \maglim\ \citep{y3-2x2maglimforecast}, defined by a magnitude cut in the $i$-band that depends linearly on the photometric redshift \citep[estimated using the algorithm of][]{de2016dnf} $z_{\rm phot}$, $i < 4z_{\rm phot} + 18$, which is split into six tomographic lens bins with redshift boundaries $[0.20, 0.40]$, $[0.40, 0.55]$, $[0.55, 0.70]$, $[ 0.85, 0.95]$, $[0.95, 1.05]$ \citep{y3-2x2ptaltlensresults}.
\end{itemize}
Angular correlation functions are evaluated in 20 log-spaced angular bins over the range $[2'.5,250']$ and then restricted by angular scale cuts (Sect.~\ref{sec:scale_cuts}). 

\begin{table}
\centering 
\begin{tabular}{|c c c|}
\hline
Parameter & Prior & Fiducial  \\ \hline
& & \\
\multicolumn{3}{|c|}{\textbf{Cosmology}} \\ 
$\Omega_{\rm{m}} $ & $\mathcal{U}[0.1, 0.9]$ & $0.3$ \\
$10^{-9}A_{\rm{s}}$ & $\mathcal{U}[0.5, 5.0]$ & $2.19$\\
 
$\Omega_{\rm{b}}$ & $\mathcal{U}[0.03, 0.07]$ & $0.048$ \\

$n_{\rm{s}}$ & $\mathcal{U}[0.87, 1.06]$ & $0.97$\\

$h$ & $\mathcal{U}[0.55, 0.91]$ &  $0.69$\\

$10^{-4} \Omega_{\nu}h^2$& $\mathcal{U}[6.0, 64.4]$ & $8.3$ \\  
$w$ & $\mathcal{U}[-2, -0.33]$ &$-1.0$\\ 

\hline
& & \\
\multicolumn{3}{|c|}{\textbf{Intrinsic Alignment}} \\ 

 $a_1$ & $\mathcal{U}[-5.0, 5.0]$ & $0.7$\\
 $a_2$ & $\mathcal{U}[-5.0, 5.0]$ & $-1.36$\\
 $\eta_1$ & $\mathcal{U}[-5.0, 5.0]$ & $-1.7$\\
 $\eta_2$ & $\mathcal{U}[-5.0, 5.0]$ & $-2.5$\\
 $b_{\rm{ta}}$ & $\mathcal{U}[0.0, 2.0]$ & $1.0$\\ 

\hline
\hline
& & \\
 \multicolumn{3}{|c|}{\redmagic\ \textbf{Galaxy Bias}} \\  
$b_1^{1\cdots 5}$  & $\mathcal{U}[0.8, 3.0]$ & $1.7,\, 1.7,\, 1.7,\, 2.0,\, 2.0$\\

\hline
 & & \\
 \multicolumn{3}{|c|}{\redmagic\ \textbf{Lens Magnification}} \\  
$C_{\rm l}^{1\cdots 5}$ & fixed & $-0.19,\, -0.63,\, -0.69,\, 1.18,\, 1.88$\\

\hline
\hline
& & \\
 \multicolumn{3}{|c|}{\maglim\ \textbf{Galaxy Bias}} \\  
$b_1^{1\cdots 6}$  & $\mathcal{U}[0.8, 3.0]$ & $1.5,\, 1.8,\, 1.8,\, 1.9,\, 2.3,\, 2.3$\\

\hline
 & & \\
 \multicolumn{3}{|c|}{\maglim\ \textbf{Lens Magnification}} \\  
 $C_{\rm l}^{1\cdots 6}$ & fixed& $0.43,\, 0.30,\, 1.75,\, 1.94,\, 1.56,\, 2.96$ \\
 \hline
\end{tabular}
\caption{The parameters varied in simulated analyses presented here, their prior ranges (using $\mathcal{U}$ to denote an uniform prior) and the fiducial values used for synthetic data. For $\Lambda$CDM analyses, $w=-1$ is fixed.}
\label{tab:params_all}
\end{table}

 \begin{figure*}
     \centering
     \includegraphics[width =0.95\textwidth]{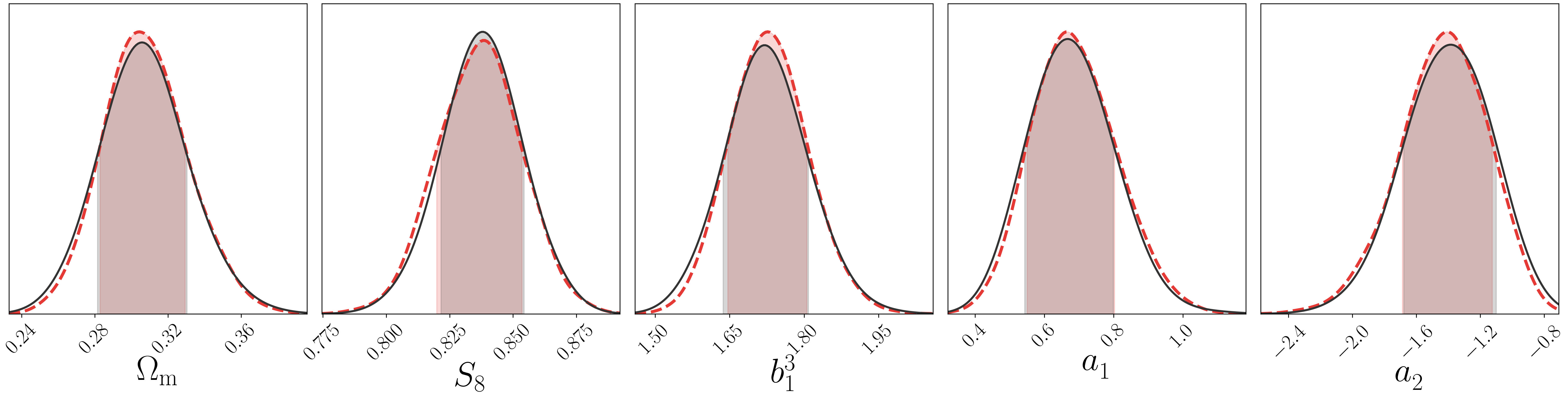}
     \caption{1D-marginalized parameter posteriors for select parameters obtained for baseline 3$\times$2pt-\redmagic\ analyses with \textsc{CosmoLike} analyzing an input data vector generated by \textsc{CosmoSIS} (black solid) and \textsc{CosmoSIS} analyzing an input data vector generated by \textsc{CosmoLike} (red dashed). The agreement of the mean posterior values demonstrates that the $\Delta \chi^2<0.2$ difference between model data vectors obtained from the two codes (at the fiducial parameter parameter point) is negligible.}
     \label{fig:code_comparison}
 \end{figure*}
We have developed two independent implementations of the baseline analysis within \textsc{CosmoSIS} and \textsc{CosmoLike}, which we compare following the procedure from DES Y1 \citep{y1methods}. These codes use different Boltzmann codes (\textsc{CAMB} or \textsc{CLASS}, respectively) and differ in terms of their structure, interpolation and integration routines. After multiple iterations we achieved an agreement of both pipelines at the level of $\Delta\chi^2<0.2$ when comparing two model vectors generated at the fiducial cosmology. A simulated likelihood analysis in the full cosmological and systematics parameter space shows no noticeable differences in the parameter constraints from both codes (c.f. Fig.~\ref{fig:code_comparison} for an illustration in a subspace of the full parameter space; here $S_8 = \sigma_8\sqrt{\Omega_{\mathrm m}/0.3}$ is a derived parameter).

\begin{figure*}
     \centering
     \includegraphics[trim= 25 40 0 200, clip, width =0.97\textwidth]{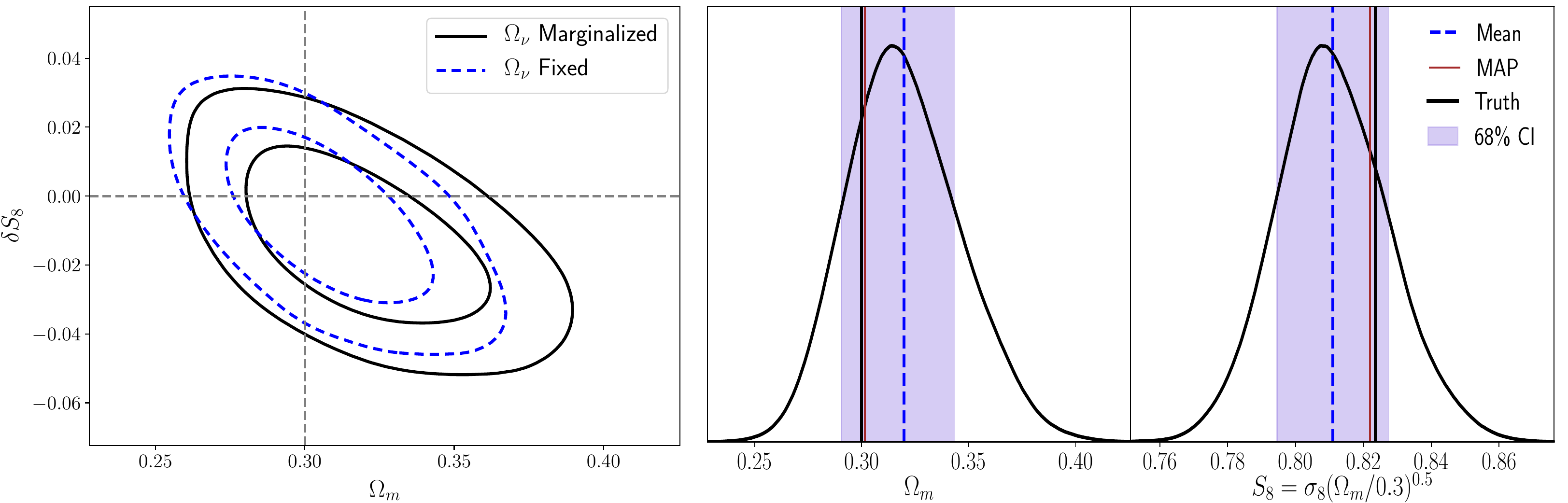}
     \caption{Marginalized 2D and 1D parameter posteriors from simulated 3$\times$2pt $\Lambda$CDM baseline analyses of synthetic, noiseless baseline data vectors.\newline\emph{Left}: Marginalized $(\Omega_{\rm m},S_8)$ posteriors assuming massless neutrinos (blue dashed) and marginalizing over neutrino mass (black solid) using the fiducial value and prior from Table~\ref{tab:params_all}. The $S_8$ posteriors are shown with respect to its input value in each analysis, which differ due to the difference in input neutrino mass, while the input matter density $\Omega_{\rm m} = \Omega_{\rm{cdm}}+\Omega_{\rm b}+\Omega_{\nu}$ is the same for both analyses (c.f.~\ref{tab:params_all}). $\Omega_\nu$ is poorly constrained by the 3$\times$2pt analysis and the shift in 2D contours indicates a projection effect from marginalizing over an under-constrained parameter that is correlated with the parameters shown. 
\emph{Center/Right}: 1D marginalized posteriors (black curve) in $\Omega_{\rm m}/S_8$ for the simulated 3$\times$2pt $\nu \Lambda$CDM baseline analysis (black solid line in the left figure). The vertical black/red lines show the input value and MAP estimate, the blue shaded bands indicate the 1D marginalized, symmetric $68\%$ uncertainty regions. The 1D marginalized mean and posterior distribution are biased from the input parameter value due to projection effect by nearly 1$\sigma$, while MAP closely recovers the input parameters.}
     \label{fig:1D_params}
 \end{figure*}

\paragraph*{Projection effects on marginalized posteriors} Even in the idealized case of applying the baseline analysis to a synthetic, noiseless data vector generated from the same model, the marginalized parameter posteriors may appear biased from the input parameter values due to parameter volume effects (c.f., Fig.~\ref{fig:1D_params}) while the maximum a posteriori point (MAP, which is optimized over the full parameter space) recovers the input parameter values. These projection effects can occur when parameters of interest are not fully constrained by the data or are degenerate with other parameters that are prior informed. We note that -- within a specific set of model and prior choices -- these projection effects decrease as the data's constraining power increases and are thus not a systematic bias. The size and direction of such projection effects depends on the parameterization and prior choices, the underlying parameter values, as well as the data's constraining power. This complicates the comparison of marginalized posteriors between different data sets and different analysis choices for the same data (e.g., variations of scale cuts, or model parameterization variations for a specific systematic effect). To indicate possible effects of parameter degeneracies on marginalized parameter constraints, 1D-marginalized DES Y3 results on a parameter $p$ are often reported as
\be
p = \text{mean value}^{+{\text{upper 34\%}}}_{-{\text{lower 34\%}}}\;\;\; (\text{MAP value})\,.
\ee

\subsection{External Data}
The most constraining results on cosmological parameters will be obtained by combining the DES 3$\times$2pt analysis with external data sets, assuming these experiments are sufficiently consistent for a combined analysis. As detailed in Sect.~\ref{sec:val}, we quantify the accuracy of model choices as bias in inferred parameters relative to the expected constraining power, which differ for each analysis (specified by data and parameter spaces). Hence we also need to validate the DES baseline model choices for the more stringent combined analysis. We simulate the constraining power of the external data at our fiducial cosmology, so that parameter biases can be attributed to model choices without any contribution from residual tension between the two experiments. For simulated joint DES-Y3+\emph{Planck} analyses, we add the constraining power of \emph{Planck} (TTTEEE+lowE) as a Gaussian prior, computed from the parameter covariance of \emph{Planck} chains \citep{Planck18_params}, to our simulated DES-Y3 analyses. This requires approximating the \emph{Planck} posterior as a Gaussian, which in the parameter dimensions relevant for DES analyses is an acceptable approximation, and shifting the \emph{Planck} best-fit value to match the fiducial parameter values from Table~\ref{tab:params_all}.

 \section{Model Validation}
 \label{sec:val}
 We scrutinize the model choices described in Sect.~\ref{sec:choices} closely following the DES-Y1 model validation procedure \citep{y1methods}, considering two different categories: 
\begin{itemize}
    \item \emph{Known, but unmodeled systematic effects}, e.g., the impact of baryons on the matter power spectrum, nonlinear galaxy bias. These systematics are mitigated through scale cuts that exclude the affected data points from the analysis.
    \item \emph{Systematics modeled with imperfect parameterizations}, which applies to all modeled systematics. We stress test the baseline parameterization to show its robustness.
\end{itemize} 

In practice, we carry out simulated (cosmic shear, 2$\times$2pt, 3$\times$2pt, 3$\times$2pt+\emph{Planck}) analyses in $\Lambda$CDM and $w$CDM on contaminated data vectors and quantify the 2D parameter bias in $(\Omega_{\mathrm m},S_8)$ for $\Lambda$CDM and in $(\Omega_{\mathrm m},w)$ for $w$CDM. We test the impact of different unmodeled systematics in combination by using contaminations at the upper limit of credible severity for each effect. For the second category, we carry out the parameterization stress test for each systematic effect individually using alternative parameterizations as input.

To ensure that the total potential systematic bias is well below $1\sigma$ statistical uncertainty, we require the 2D parameter biases to be smaller than $0.3\sigma_{2\mathrm{D}}$; additionally we require a residual $\Delta \chi^2<1$ (after fitting the baseline model to the contaminated data) for each of these tests in order to not bias the goodness of fit. If a model variation changes the data vector by $\Delta \chi^2 < 0.2$ \emph{without refitting}, we consider this model variation to be insignificant as the change in $\Delta \chi^2$ is less than the residual between the two analysis codes. If a model variation changes the data vector by  $\Delta \chi^2 < 1$ without refitting, the $\Delta \chi^2$ threshold after refitting is automatically met; hence we use importance sampling (IS) of the baseline chain to estimate parameter biases, which is computationally more efficient than carrying out an independent analysis.

We emphasize that all model validation tests in this paper were carried out for both lens samples in parallel, and both lens samples pass these validation criteria. For clarity of presentation, we primarily describe quantitative results for the \redmagic\ sample. Corresponding figures and numbers for the \maglim\ sample can be found in Ref.~\citep{y3-2x2ptaltlensresults}.

When small differences in likelihood correspond to a large region in parameter space, e.g., for insufficient or degenerate models, we found MAP parameter estimates to be noisy even with repeated numerical optimization using the Nelder-Mead method \citep{NeldMead65} starting from the best-fit parameter values obtained from likelihood chain.
Hence we determine parameter biases from marginalized 2D-parameter posteriors. To account for projection effects, we evaluate parameter biases relative to the marginalized 2D-parameter posteriors of an otherwise identical analysis of the uncontaminated data vector, which is subject to the same projection effects (cf. Fig~\ref{fig:scale_cuts_LCDM}). While the size and direction of parameter projection effects can be altered by large changes in the input parameters, we verified their stability with respect to small fluctuations by analyzing 100 noisy data vector realizations with noise drawn from the data covariance. 
\subsection{Scale Cuts}
\label{sec:scale_cuts}
For the DES-Y3 baseline model, the largest unmodeled systematics are the impact of baryonic feedback on the matter power spectrum and nonlinear galaxy bias. As both of these effects will affect the 3$\times$2pt data, they need to be mitigated in combination. However, in practice cosmic shear scale cuts are driven by baryonic feedback modifying the matter power spectrum, while nonlinear galaxy biasing is the dominant contamination to galaxy clustering and galaxy-galaxy lensing.

We optimize scale cuts through a series of simulated analyses of a synthetic data vector contaminated by baryonic feedback effects and nonlinear galaxy bias, using different scale cut proposals. If the contamination model assumes a realistic upper boundary for the effect, the analysis will yield unbiased parameter values.
The theoretical model error covariance \citep{Baldauf2016} provides an elegant alternative to scale cuts through analytic marginalization of theoretical model uncertainties. However, this approach requires a model for the distribution of the model error, an assumption that is much more difficult to validate than the choice of one pessimistic contamination realization for scale cuts. It was recently demonstrated \citep{Chudaykin2020} that the model error covariance for redshift-space galaxy power spectrum analyses can be obtained from large mock catalogs. However, an application to baryonic feedback effects requires additional research, such as how to fairly sample the space of possible feedback models\footnote{A first exploratory work to incorporate baryonic effects using a theoretical error covariance without a rigorous statistical justification was recently performed 
\citep{2021arXiv210401397M} with promising results.}.
 \subsubsection{Baryon Impact Modeling}
 We bracket the impact of baryonic physics on the DES data vector using measurements of the matter power spectrum from hydrodynamic simulations with large AGN feedback. As in the DES-Y1 analysis \citep{y1methods,y1kp}, we use the \textsc{AGN} scenario from the OWLS simulation suite \citep{OWLS,dsb11}, which adopts the AGN subgrid physics under the prescription of Ref.~\citep{BoothAGN}; black holes inject 1.5\% of the rest mass energy of the accreted gas into the surrounding matter in the form of heat. 
 
We consider \textsc{OWLS-AGN} an adequate upper limit for the impact of baryonic physics. \citet{hem2020} ranked various hydrodynamical scenarios based on the amount of suppression toward small scales against the dark-matter-only theoretical prediction in the cosmic shear statistics (see their Figs. 15 and 16).
While more extreme scenarios, most notably the \textsc{Illustris} simulation \citep{Vogelsberger2014}, exist, \textsc{Illustris} is known be too extreme in its radio-mode AGN feedback, and underpredicts the amount of baryons in galaxy groups compared with observations (see Fig.1 of \citealt{Haider16} or Fig.10 of \citealt{Genel14}). The successor \textsc{IllustrisTNG} simulation \citep[e.g., see][]{TNG1, TNG2} modifies the original \textsc{Illustris} feedback prescription, resulting in much weaker suppression effect in the summary statistics of power spectrum (see Fig.1 of Ref.~\citep{hem2020}). 
The baseline \textsc{BAHAMAS} simulation \citep{bahamas} is calibrated with observations of the present-day stellar mass function and hot gas fractions in galaxy groups to ensure the overall matter distribution are broadly correct under the effect of baryons. It improves the subgrid models of the \textsc{OWLS-AGN} simulation, which has overly efficient stellar feedback and underpredicts the abundance of $<10^{11} {\rm M}_{\odot}$ galaxies at present day \citep{McCarthy17}. When performing validation, we thus select the scenario of \textsc{OWLS-AGN} which is more extreme than the \textsc{BAHAMAS}, but below the suppression level of \textsc{Illustris}\footnote{The high AGN feedback version of \textsc{BAHAMAS} (tagged as BAHAMAS T8.0 in \citealt{hem2020}) exhibits feedback strengths between the default \textsc{BAHAMAS} and \textsc{Illustris}; the amount of suppression in cosmic shear is similar to \textsc{OWLS-AGN}.}.

We refer to \cite{hem2019,hem2020} for numerical details on the computation of contaminated DES model data vectors.
 
 \subsubsection{Nonlinear Galaxy Bias Modeling}
 We model the contribution of nonlinear galaxy biasing to galaxy clustering and galaxy-galaxy lensing using an effective 1-loop model with renormalized nonlinear bias parameters \citep{McDonaldRoy,baldauf12,chan12,Saito_bnl}, $b_2$ (local quadratic bias), $ b_{s^2} $ (tidal quadratic bias) and $ b_{\rm 3nl} $ (third-order nonlocal bias):
 \begin{align}\label{eq:P_tm}
\nonumber P_{g\mathrm{m}}(k)=&\; b_1 P_{\mathrm{mm}}(k) +  \frac{1}{2} b_2P_{ b_1 b_2}(k) + \frac{1}{2} b_{s^2}P_{\rm b_1 s^2}(k)\\ 
&+ \frac{1}{2} b_{\rm 3nl}P_{ b_1 b_{\rm 3nl} }(k)\,,\\
\nonumber	P_{gg}(k) =&\; b_1^2 P_{\mathrm{mm}}(k) + b_1b_2 P_{\rm b_1 b_2}(k) + b_1b_{s^2}P_{ b_1 s^2}(k)\\
\nonumber & +  b_1b_{\rm 3nl}P_{ b_1 b_{\rm 3nl} }(k) +  \frac{1}{4}b_2^2 P_{ b_2 b_2}(k)\\
& + \frac{1}{2}b_2b_{s^2}P_{ b_2 s^2}(k) + \frac{1}{4}b_{s}^2 P_{ s^2 s^2}(k)\,, 
\end{align}
where we have omitted the subscript denoting the dependence of bias parameters on the galaxy sample (e.g., $b_{2,g})$ for clarity. Expressions for the power spectrum kernels $P_{b_1 b_2},$ etc., are given in \citet{Saito_bnl}. 

\begin{figure}
    \centering
    \includegraphics[width= 0.49\textwidth,trim=0.6cm 2.1cm 0cm 1.6cm, clip]{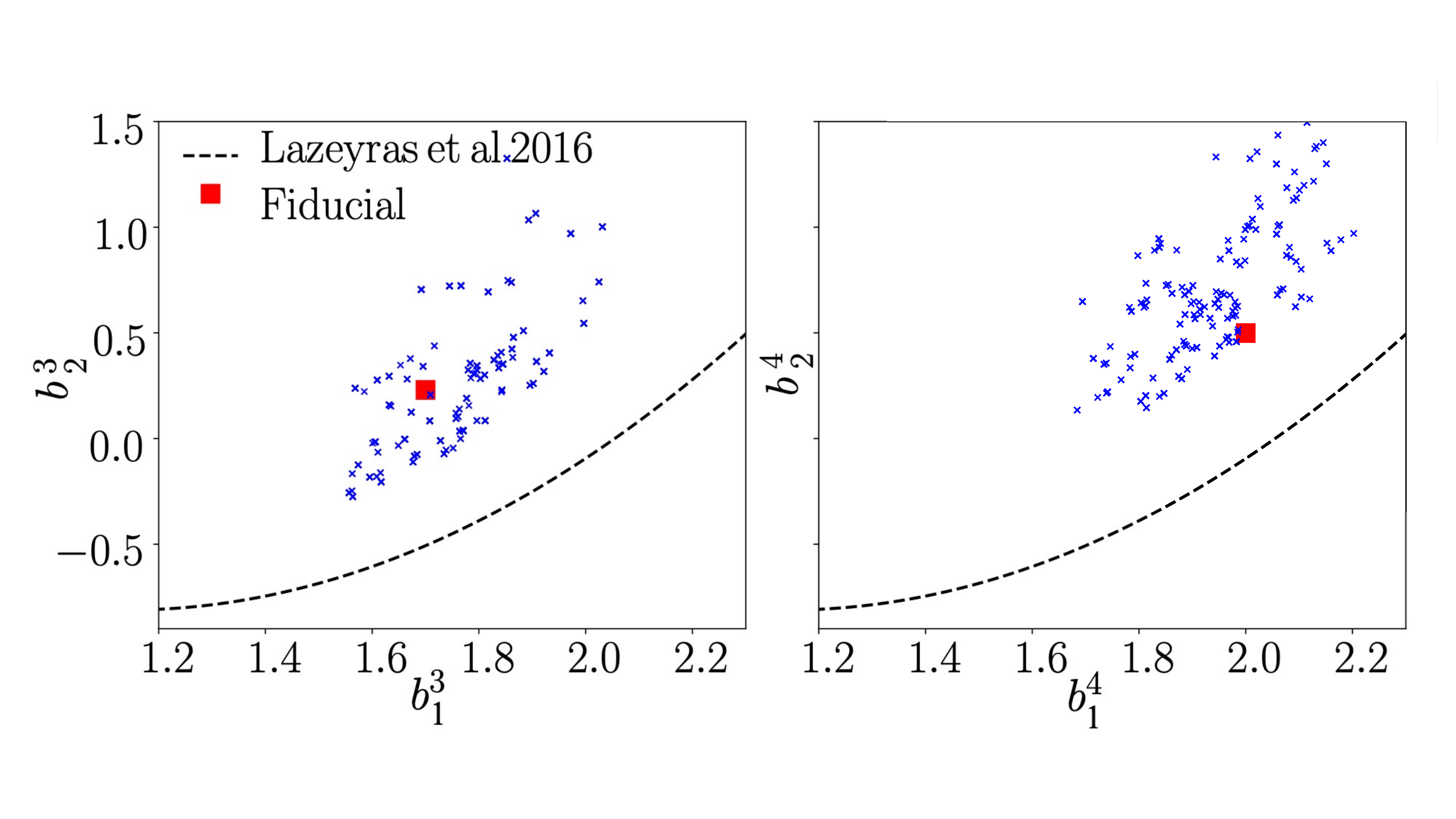}
    \caption{Relation between linear bias $b_1$ and local quadratic bias $b_2$ for halos \citep[dashed line]{Lazeyras16} and galaxies (blue crosses), obtained by averaging the halo relation over HODs constrained by the \redmagic\ abundance and linear bias values (blue crosses), for tomographic lens bins 3 (left) and 4 (right). The red squares denote the values used to compute the nonlinear bias contamination in Sect.~\ref{sec:scale_cuts}.}
    \label{fig:b2_b1}
\end{figure}

This model was found to describe the clustering of \redmagic-like galaxies in mock catalog at DES-Y3 accuracy down to 4 $\rm{Mpc}/h$ \citep{bias3D}. As validated in \cite{bias3D}, we fix the bias parameters $b_{s^2}$ and $ b_{\rm 3nl} $ to their co-evolution value of $b_{s^2}=-4(b_1 - 1)/7$ and $b_{\rm 3nl}=b_1 - 1$ \citep{Saito_bnl}. For $b_2$, we use values interpolated from the $b_2-b_1$ relation measured from mock catalogs of \redmagic-like galaxies \citep{bias3D}, $b_2(b_1=1.7) = 0.23,\,b_2(b_1=2.0) = 0.5 $. For comparison, $b_{2,\rm{halo}}(b_{1,\rm{halo}}=1.7) = -0.51,\,b_{2,\rm{halo}}(b_{1,\rm{halo}}=2.0) = -0.09 $ using the fitting function of \citet{Lazeyras16} for halos. To motivate our choice for $b_2$, we calculate the impact of broad halo occupation distributions (HOD) by averaging $b_{2,\rm{halo}}(b_{1,\rm{halo}})$ over HODs that are only constrained by the \redmagic\ abundance and linear bias values (c.f. discussion of galaxy bias in Sect.~\ref{sec:val} for details on these HODs). Figure~\ref{fig:b2_b1} shows that the $b_2$ values measured in mocks are consistent with those expected for galaxy samples with a broad halo mass distribution.
\subsubsection{Scale Cut Analyses}
\begin{figure*}
    \centering
    \includegraphics[width=0.338\textwidth,trim=0.4cm 0.3cm 2.0cm 0.2cm, clip]{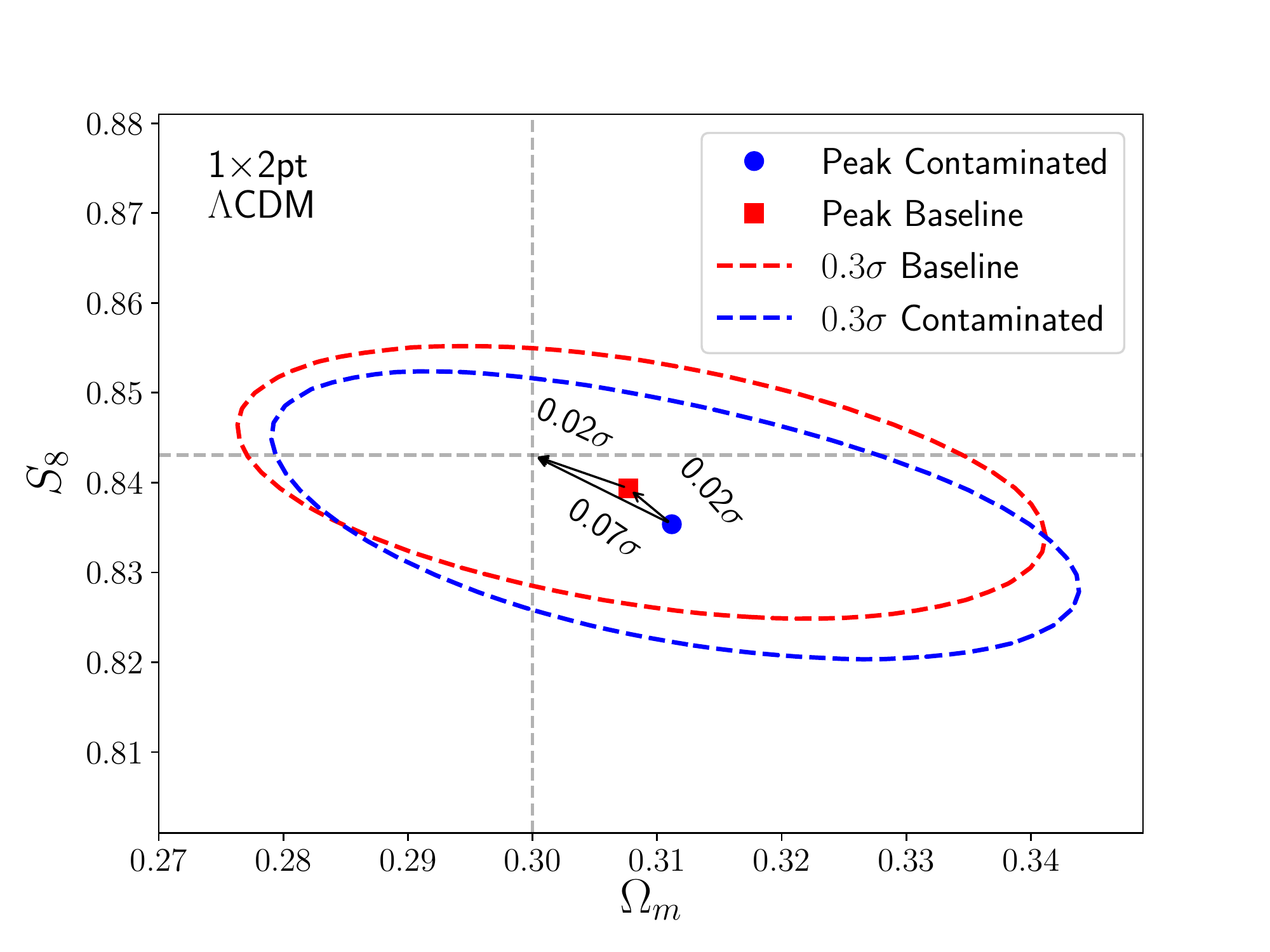}
    \includegraphics[width=0.3\textwidth,trim=2.54cm 6.1cm 2.04cm 7.2cm, clip]{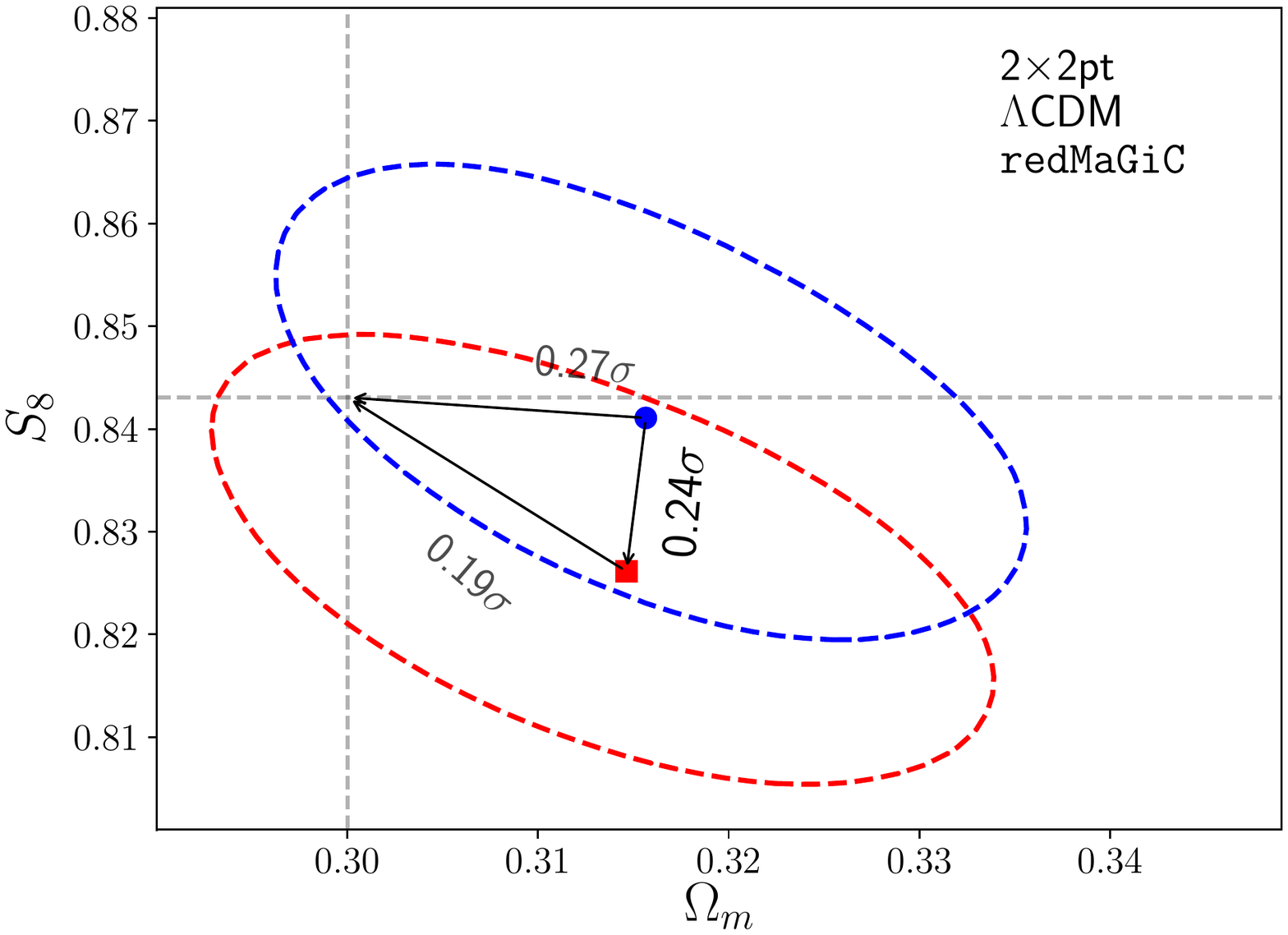}
    \includegraphics[width=0.3\textwidth,trim=2.54cm 5.85cm 1.85cm 7.2cm, clip]{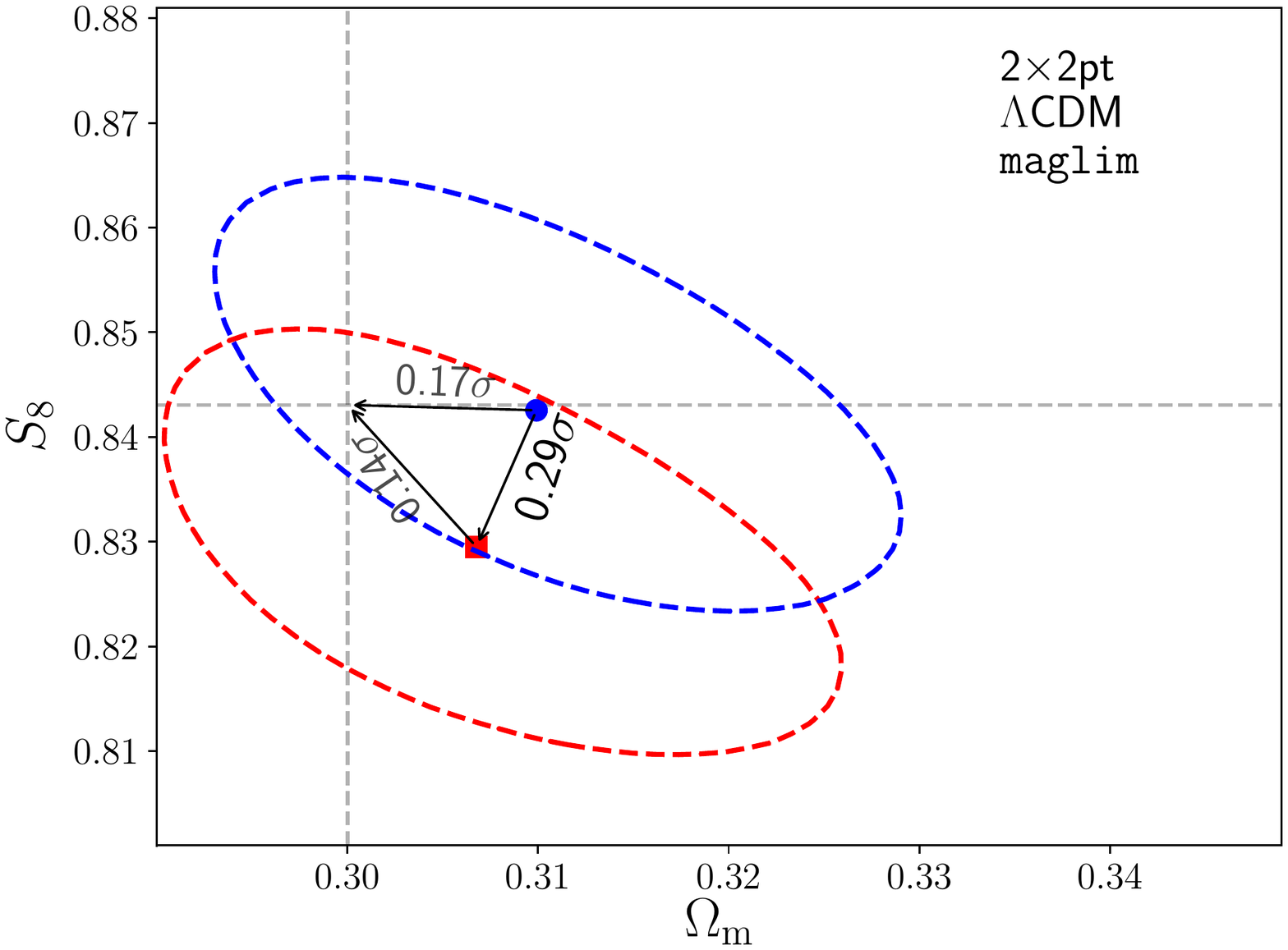}
    \caption{Parameter biases for final scale cuts in $\Lambda\mathrm{CDM}$: the red and blue ellipses show $0.3\,\sigma$ contours for the 2D marginalized constraints of the baseline and contaminated data vector, respectively. Due to parameter volume effects, the marginalized constraints from the baseline analysis are not centered on the input cosmology, and the parameter bias due to unaccounted systematics is given by the offset between the red and blue contours. \emph{Left:} Simulated cosmic shear analyses. \emph{Center:} Simulated 2$\times$2pt-\redmagic\ analyses. \emph{Right:} Simulated 2$\times$2pt-\maglim\ analyses. The dashed horizontal and vertical lines indicate the input parameter values.}
    \label{fig:scale_cuts_LCDM}
\end{figure*}
\begin{figure*}
    \centering
    \includegraphics[width=0.34\textwidth,trim=0.4cm 6.1cm 2.0cm 7.2cm, clip]{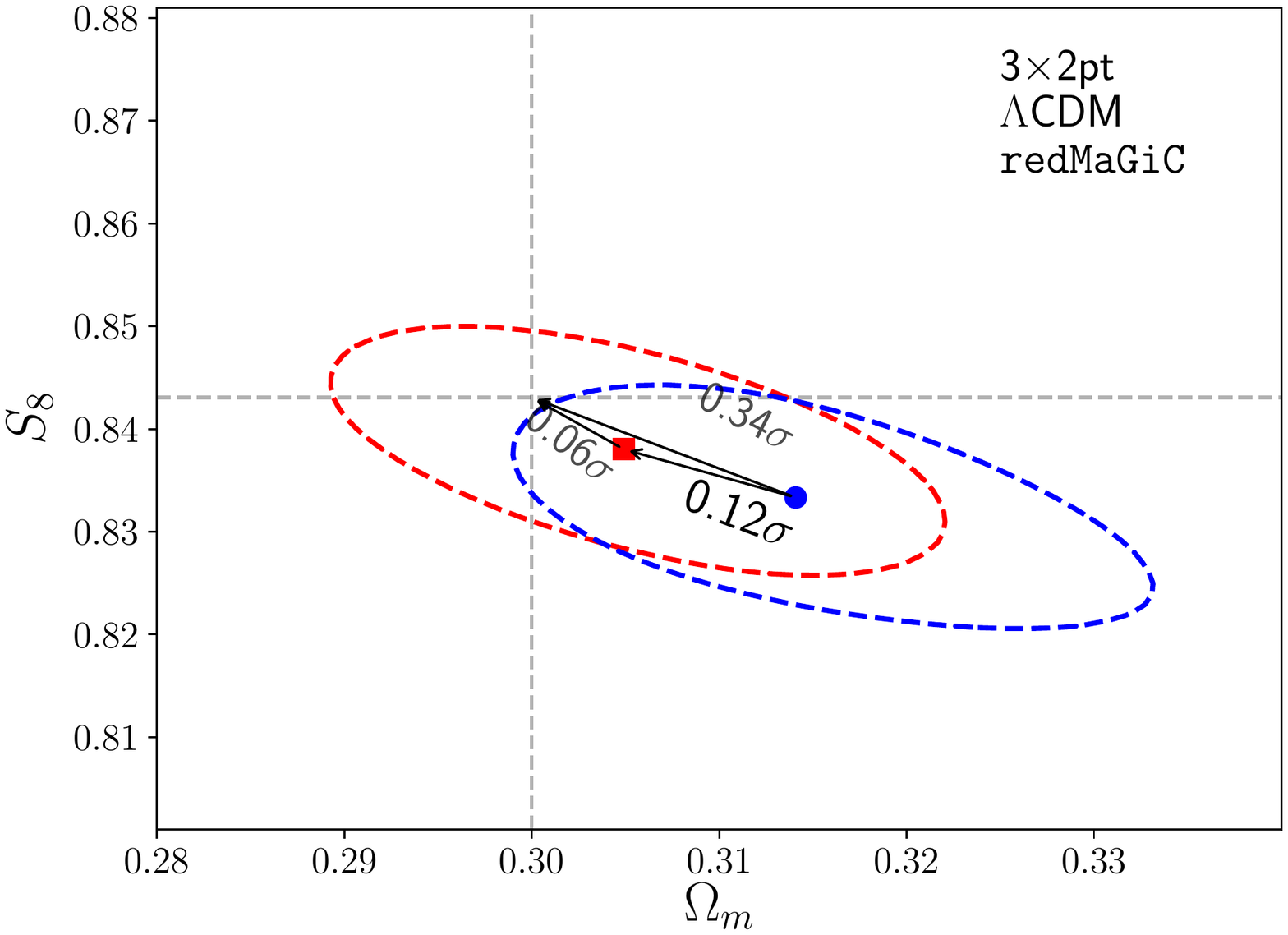}
    \includegraphics[width=0.3\textwidth,trim=2.58cm  6cm 2.0cm 7.2cm, clip]{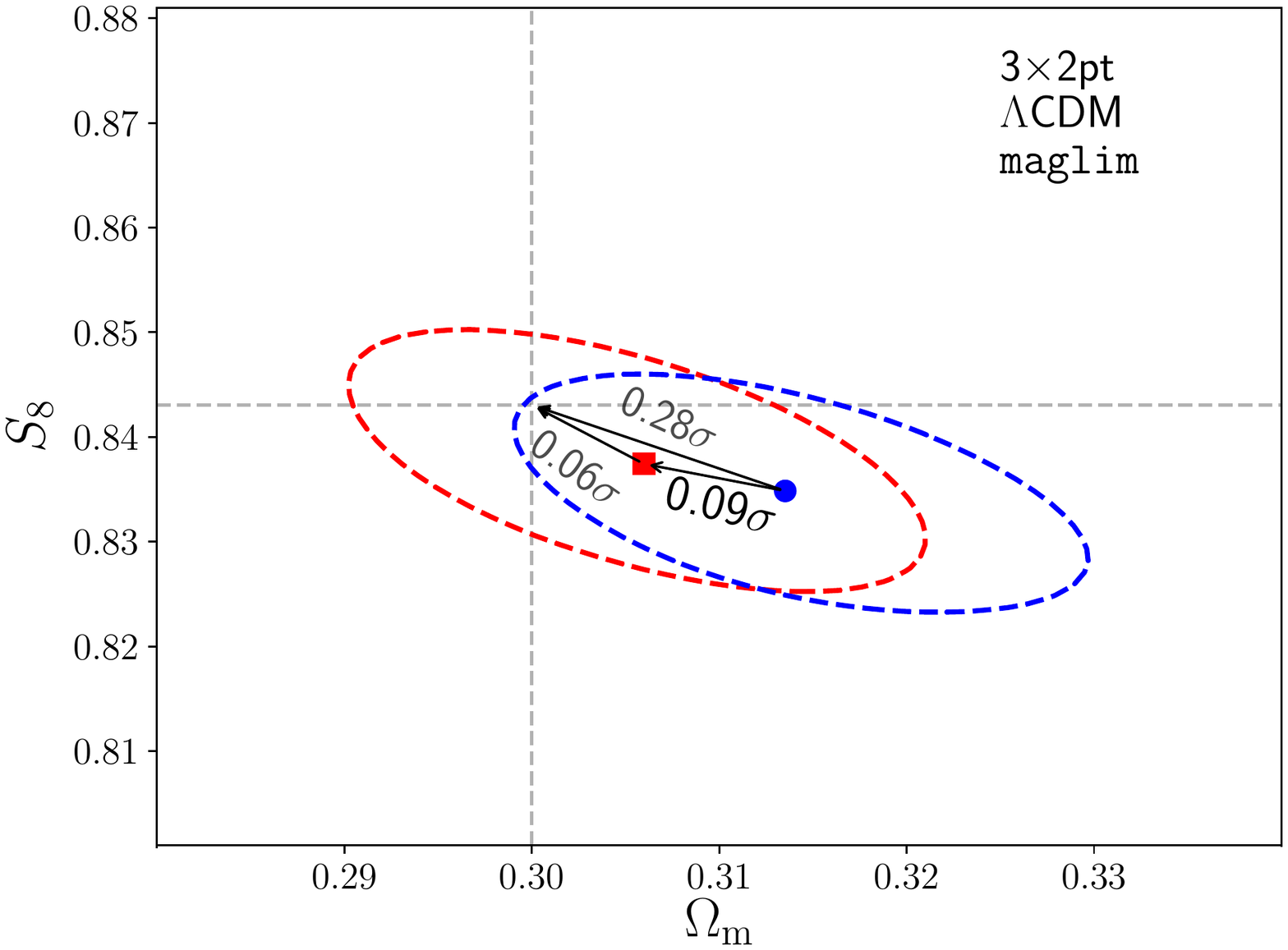}
    \includegraphics[width=0.33\textwidth,trim=0.53cm 0 2.0cm 0.2cm, clip]{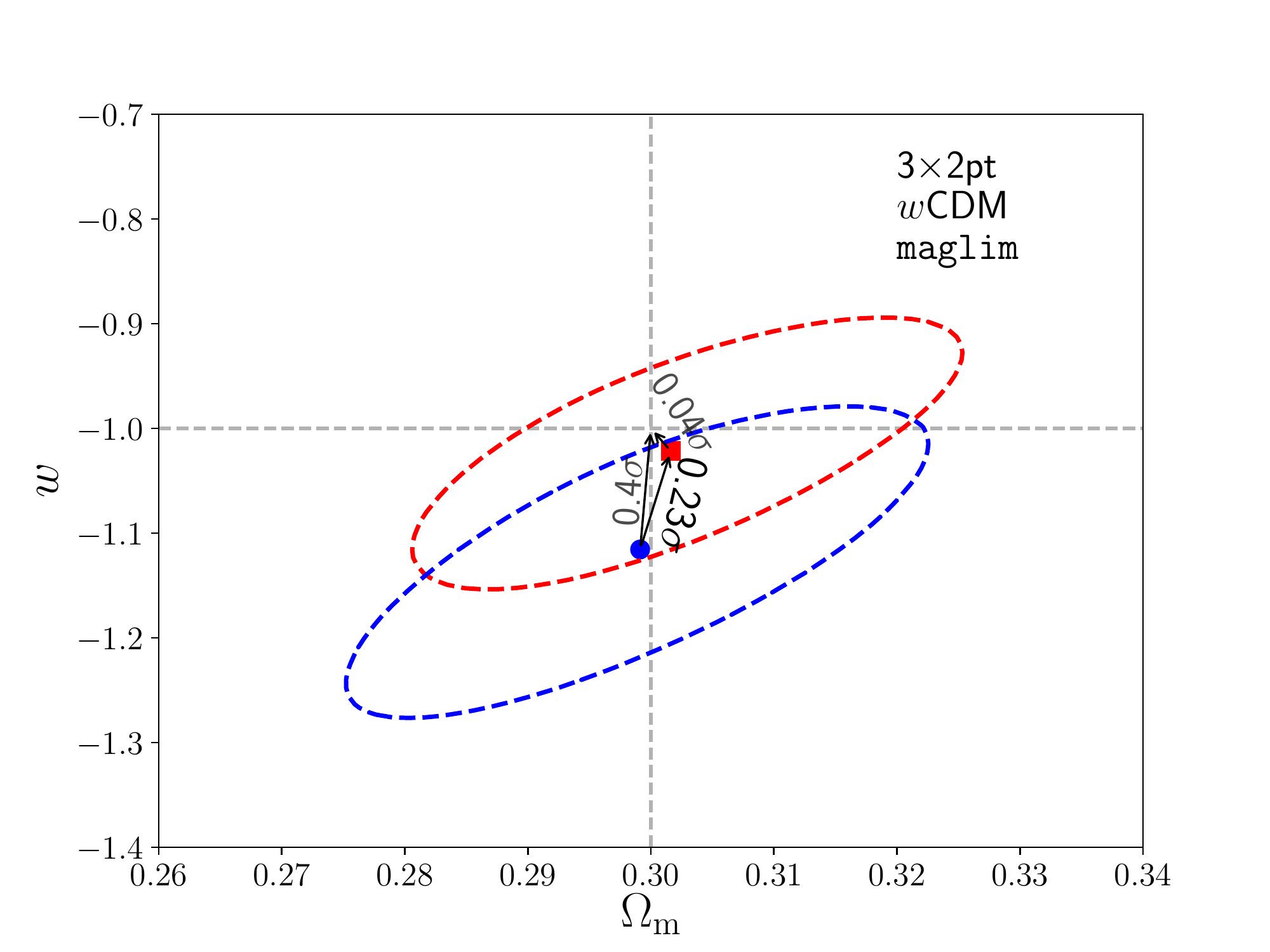}
    \caption{Same as Fig.~\ref{fig:scale_cuts_LCDM} for 3$\times$2pt-analyses in  $\Lambda\mathrm{CDM}$ and $w\mathrm{CDM}$ analyses. \emph{Left:} Simulated $\Lambda\mathrm{CDM}$ 3$\times$2pt-\redmagic\ analyses. \emph{Center:} Simulated $\Lambda\mathrm{CDM}$ 3$\times$2pt-\maglim\ analyses. \emph{Right:} Simulated $w\mathrm{CDM}$ 3$\times$2pt-\maglim\ analyses.}
    \label{fig:scale_cuts_wCDM}
\end{figure*}
We generate data vectors with the baryonic feedback and nonlinear bias contamination models described above, and run simulated likelihood analyses for a family of angular scale cut proposals: for galaxy clustering and galaxy-galaxy lensing, we vary the minimum comoving transverse scale $R_{\rm{min}}$ included in the analysis, corresponding to an angular scale cut
$
\theta_{\mathrm{min}}^i  =R_{\mathrm{min},w/\gamma_{\rm{t}}}/\chi(z^i)
$
for each tomographic lens bin $i$.

For cosmic shear, we choose a threshold $\Delta\chi^2_{\rm{thr}}$ value for the difference between contaminated and baseline cosmic shear data vectors, 
and then determine scale cuts $\theta_{\rm{min},\pm}^{ij}$ for each pair of source bins such that the baryonic contamination results in $\Delta\chi^2(\xi_{\pm}^{ij})<\Delta\chi^2_{\rm{thr}}/20$ for each of the 20 shear 2pt functions (10 different tomographic combinations each for $\xi_+$ and $\xi_-$).

While scale cuts need to be tested for each analysis setup (choice of probes, model space, and parameters of interest), we require the same scale cuts for cosmic shear, 2$\times$2pt, 3$\times$2pt and 3$\times$2pt+\emph{Planck} in $\Lambda$CDM and $w$CDM analyses in order to enable model comparisons based on the same data points. We find that scale cuts corresponding to the parameters 
\be
\left(R_{\mathrm{min},w},\,R_{\mathrm{min},\gamma_{\rm{t}}},\Delta\chi^2_{\rm{thr}}\right) =\big(8\, \mathrm{Mpc}/h,\,6\, \mathrm{Mpc}/h,\, 0.5\big)\,
\ee
meet our parameter bias requirements for all these analyses.
Figure~\ref{fig:scale_cuts_LCDM} illustrates the residual $\Lambda$CDM parameter biases in cosmic shear, 2$\times$2pt and 3$\times$2pt. The corresponding 2D parameter bias for 3$\times$2pt+\emph{Planck} ($\Lambda$CDM) is $0.02\sigma_{\mathrm{2D}}$; for $w$CDM we found parameter biases of $0.1\sigma_{\mathrm{2D}}$, $0.16\sigma_{\mathrm{2D}}$ and $0.05\sigma_{\mathrm{2D}}$ for 2$\times$2pt, 3$\times$2pt, and 3$\times$2pt+\emph{Planck} respectively.

At these fiducial scale cuts, even an extreme contamination such as the Illustris feedback model only results in a $0.23\sigma_{\mathrm{2D}}$ and $0.50\sigma_{\mathrm{2D}}$ for shear and 3$\times$2pt ($\Lambda$CDM), respectively, which demonstrates the robustness of our analysis choices.

In addition to the fiducial scale cuts described above, \citet{y3-cosmicshear1,y3-cosmicshear2} introduce scale cuts for an ``optimized'' cosmic shear analysis, which maximizes the constraining power of cosmic shear, such that $\sigma_{2\mathrm{D},3\times2\text{pt}} = 0.3$ in $\Lambda$CDM. For the cosmic shear analysis, these optimized scale cuts result in $\sigma_{2\mathrm{D},\text{1x2pt}} = 0.1$. The corresponding scale cuts were obtained by iteratively removing the data point from the tomographic cosmic shear data vector with the largest contribution to $\Delta \chi^2$, until the baryonic contamination results in a residual $\Delta\chi^2 =1$. $\Lambda$CDM constraints from a 3$\times$2pt analysis  based on these optimized cosmic shear scale cuts are presented in Ref.~\citep{y3-3x2ptkp}. We note that this scale cut choice does not pass the validation criteria in $w$CDM and that it was introduced after unblinding of the cosmology results.
 \subsection{Model Parameterization Stress Tests}
 \label{sec:stress_tests}
 \subsubsection{Matter Power Spectrum}
In order to bracket potential parameter biases due to the accuracy limitations of \textsc{halofit} for the nonlinear gravity-only power spectrum, we compare to the  \textsc{Cosmic Emu} \citep{emu_lawrence} and \textsc{Euclid Emulator} \citep{emu_euclid} emulators, which are based on larger, more recent simulations and thus provide higher model accuracy. These more accurate emulators are designed for a limited range in cosmological parameter space, which does not cover the expected $2\sigma$ parameter uncertainty regions of DES-Y3 cosmic shear and 2$\times$2pt analyses. Hence current emulators are well suited for validation at selected cosmologies, but are not a practicable alternative for the baseline model.

\begin{figure*}
     \centering
     \includegraphics[width=0.98\textwidth]{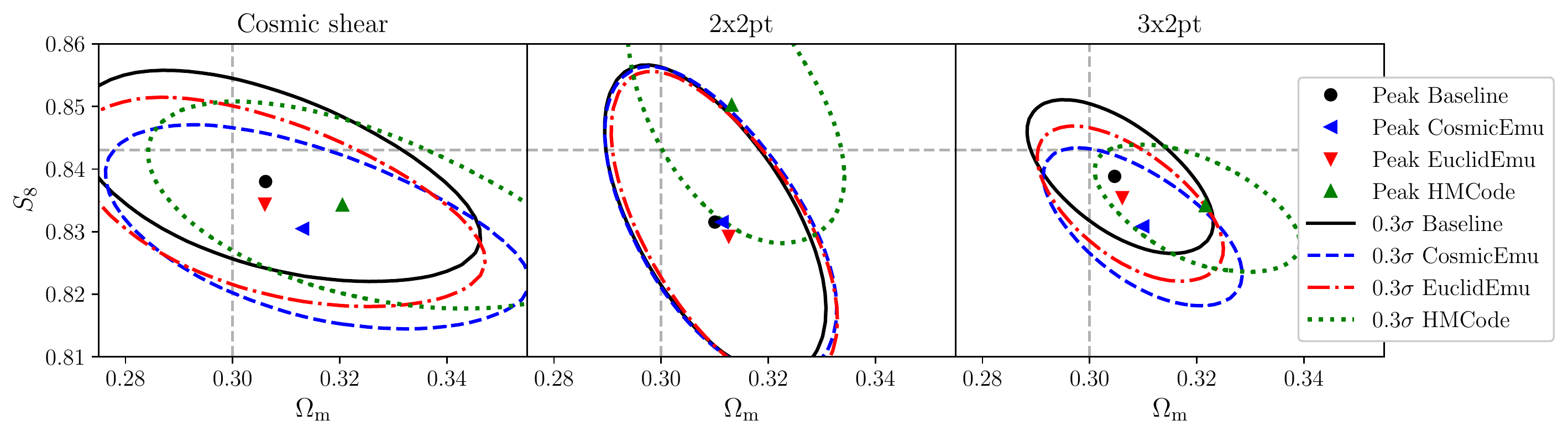}
     \caption{Robustness of parameter constraints to the choice of matter power spectrum model. The black contour shows the $0.3\sigma_{2\rm{D}}$ baseline analysis results (analyzing \textsc{halofit} input with \textsc{halofit} model). The blue/red/green contours illustrate the systematic bias, estimated through importance sampling, when input data from \textsc{Cosmic Emu/Euclid Emulator/HMCode} is analyzed with \textsc{halofit}. All contours levels are $0.3\sigma_{2\rm{D}}$, and symbols indicate the 2D (marginalized) peak.}
     \label{fig:test_mm}
 \end{figure*}

For completeness, we also compare to the \textsc{HMCode} fitting function \citep{HMCode15}. Figure~\ref{fig:test_mm} shows the result of analyzing simulated data vectors computed from these alternate prescriptions for the nonlinear gravity-only power spectrum with \textsc{halofit}. We find insignificant parameter biases relative to the two emulator models ($<0.15\sigma_{2\rm{D}}$). We note that the comparison of \textsc{HMCode} and \textsc{halofit} fails our requirements for the 2$\times$2pt analysis. The \citet{HMCode15} version of \textsc{HMCode} employed here is known to overpredict the power spectrum in the quasilinear regime ($k\sim 0.1\dots0.5 h/\mathrm{Mpc}$, \citep[c.f. Fig.~1 of][]{KIDS_methods}); the DES 2$\times$2pt analysis is particularly sensitive to these scale due to the clustering scale cut $R_{\mathrm{min},w} =8 \mathrm{Mpc/h}$. While these comparisons indicate that \textsc{halofit} is sufficient for the DES-Y3 baseline analysis, the accuracy requirements of future weak lensing surveys on matter power spectrum modeling \citep[e.g.,][]{euclid_nl} may require a new generation of models/emulators with sufficient support in parameters.
\subsubsection{Galaxy Bias}
  \begin{figure}
     \centering
     \includegraphics[width=0.4\textwidth]{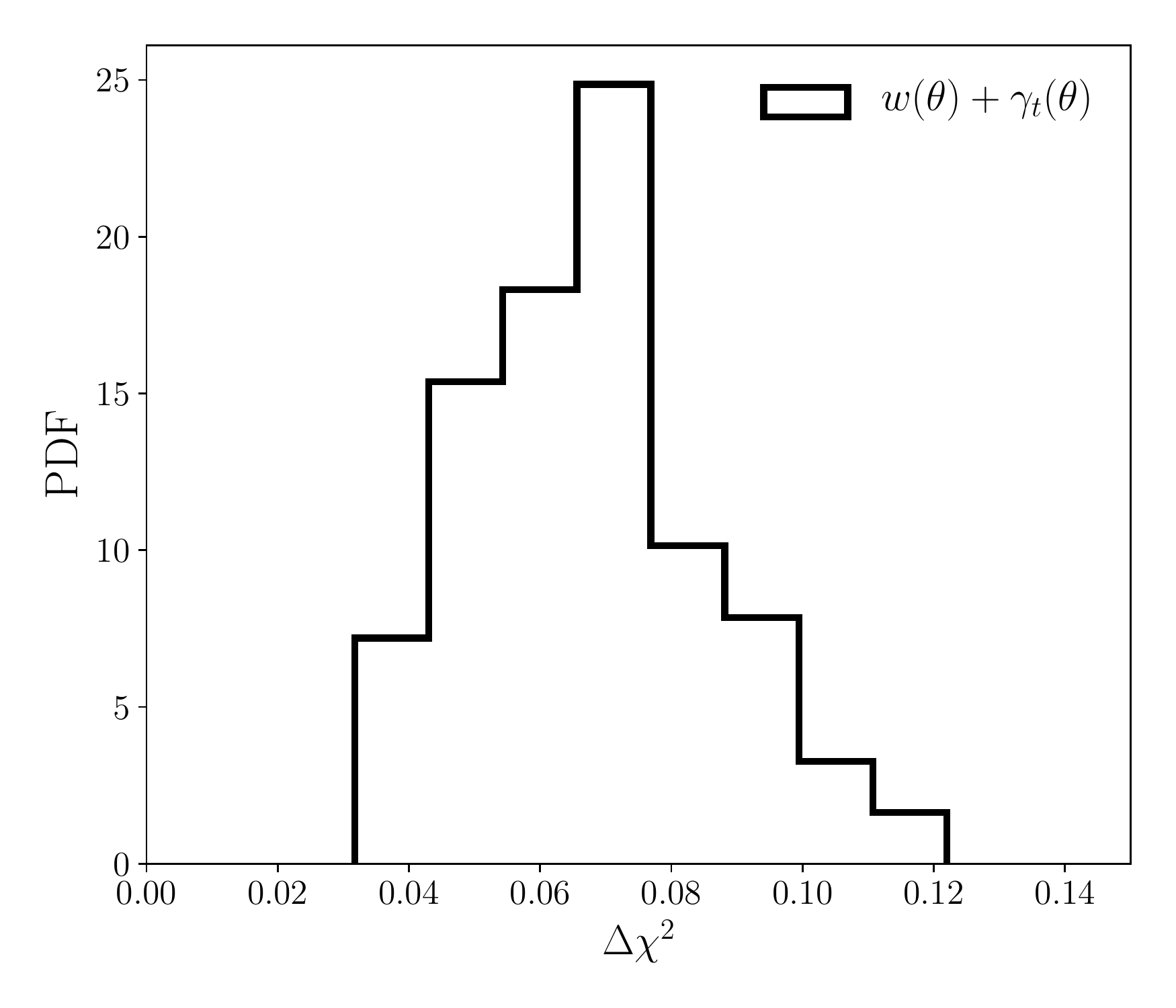}
     \caption{Impact of galaxy bias redshift evolution on DES 2pt statistics: We compute $\Delta \chi^2$ between models with galaxy bias constant within each tomographic bins, and models with evolving galaxy bias based on the redshift evolution of the halo mass function and halo mass--bias relation, assuming a constant HOD within each tomographic bin. The histogram shows the $\Delta \chi^2$ distribution when sampling HOD parameters that are constrained only by the fiducial linear bias value and observed galaxy number density.}
     \label{fig:test_b}
 \end{figure}
The scale cuts derived in Sect.~\ref{sec:scale_cuts} are designed to enable a linear galaxy bias baseline model. It remains, then, to validate the assumption that these linear bias parameters can be modeled as constant within each tomographic bin. Two independent mock catalogs \citep{buzzard,MICE_HOD} display limited redshift evolution of the HOD of \redmagic\ galaxies within the relatively narrow ($\Delta z = 0.15$) tomographic bins. Hence we model the potential redshift evolution of galaxy bias within tomographic bins based on the redshift evolution of the halo mass function and halo mass--halo bias relation assuming a constant HOD. For each tomographic bin, we generate samples of possible HODs using the parameterization from Ref.~\citep{Zheng_HOD} for luminosity threshold samples, with an additional parameter to account for the incompleteness of central galaxies in a color-selected sample, constrained by the fiducial linear bias values and observed number density \citep[see][for details]{y3-2x2ptbiasmodelling}. Figure~\ref{fig:test_b} shows that this model for bias redshift evolution results in $\Delta \chi^2 < 0.2$. Hence we expect the redshift evolution of galaxy bias to have negligible impact on parameter constraints, even if one allowed for redshift evolution of the HOD within $\Delta z=0.15$ tomographic bins.

We also tested that the scale dependence of galaxy bias imparted by massive neutrinos \citep{LoVerde14} has no significant impact on inferred parameters.
\subsubsection{Intrinsic Alignments}
The TATT model adopted as the baseline intrinsic alignment parameterization includes two next-to-leading order effects (density weighting, tidal torquing) which are theoretically well-motivated but observationally less well studied. \citet{Samuroff_2019} found an indication for nonzero values of the associated parameters ($b_\mathrm{ta}, a_2$) at around the $1\sigma$ level in cosmic shear alone, with the less constraining DES-Y1 data. We thus include these parameters in the fiducial setup, which was frozen before unblinding of the DES-Y3 analysis, noting that their redshift evolution has not yet been constrained. Also before unblinding, it was agreed to characterize their redshift evolution in post-unblinding work \citep{y3-cosmicshear2} if the DES-Y3 analysis found these types of intrinsic alignment effects to be significant.

While perturbative IA models that are more complete and extend to higher order than TATT exist \citep{Vlah_2020}, there are no observational constraints on the additional parameters. 
Due to the lack of realistic amplitudes for higher-order IA contaminations, we do not use a strategy analogous to the perturbative bias contamination in Sect.~\ref{sec:scale_cuts} to test whether TATT is sufficient to model the scale-dependence of IA in the context of this analysis.
Furthermore, we do not test robustness to fully nonlinear IA contributions, noting that \citep{fortuna2020} found the 1-halo IA contamination unlikely to significantly bias current generation lensing studies.
 
The redshift and galaxy luminosity dependence of NLA have been characterized (for elliptical galaxies) over limited ranges in luminosity and redshift \citep{MegaZ,lowz}. Assuming that only red galaxies are subject to linear alignments, one can predict the mean NLA amplitude of the DES source sample as a function of redshift through extrapolation (in luminosity and redshift) of the observed NLA scalings down to the limiting magnitude over the redshift range \citep{krause16}. This estimate further requires extrapolation of an observed luminosity function for all and red galaxies in order to model the red fraction of galaxies \citep{krause16}. The redshift evolution predicted from extrapolation of the NLA amplitude of the MegaZ-Luminous Red Galaxy sample and the DEEP2 luminosity functions \citep{DEEP2} is not monotonic and not well described by a power law. This prescription was used in DES Y1 \citep{y1methods} to demonstrate that a power law parameterization for the NLA redshift evolution was sufficient for the DES-Y1 analysis, and predicted an NLA amplitude for the DES-Y1 source sample $A_1(z_0)$ consistent with the constraints obtained in Ref.~\citep{y1kp}.

For DES-Y3, this model for the redshift dependence of $A_1$ results in parameter shifts up to $0.5\sigma_{2\rm{D}}$ for cosmic shear alone in the context of the TATT model with fiducial $a_2,\, \eta_2,\,b_\mathrm{ta}$ value.
It does not result in significant biases for 2$\times$2pt or 3$\times$2pt analyses and passes our requirements, except for the cosmic shear analysis. Hence we validate the robustness of this parameterization \emph{a posteriori} on data. This \emph{a posteriori} validation procedure, as well as several additional IA model robustness tests for cosmic shear analyses, are presented in \citet*{y3-cosmicshear2}. We note that the realism of this model stress test is questionable as it is based on several extrapolations ($z-$ and $L-$ scaling of observed $A_1$; $z$, $L$ dependence of the luminosity function for all and red galaxies). Due to the absence of a sharp test or other well-motivated parameterizations, we proceed with the baseline power-law redshift evolution parameterization and  \emph{a posteriori} validation on data.
 
\subsubsection{Magnification}
\citet{y3-2x2ptmagnification} demonstrate that DES-Y3 cosmology constraints are robust to biases in the estimated values for $C_{\rm{l}}^i$, including the extreme scenario of $C_{\rm{l}}^i=0$, i.e., ignoring lens magnification in the analysis. Hence we leave tests of the redshift evolution of these coefficients to future analyses, for which lens magnification is forecasted to become a significant systematic \citep{Lorenz_mag,Thiele_mag}. 

 \subsubsection{Higher-Order Lensing Effects}
\label{sec:test_g}
 The calculations of angular statistics described in Sect.~\ref{sec:theory} included only leading-order lensing effects. The most significant next-to-leading-order contributions to the observed two-point statistics are reduced shear \citep{Dodelson06,Shapiro09} and source clustering and magnification \citep{Peter_Bmodes,lensingbias}, while distortions from multiple deflections can safely be ignored \citep{CH02,KH10}. We review these effects below, and derive expressions for the impact of source clustering applicable to broad tomographic source bins.
 
Each of these effects amounts to weighting the shear by another field (convergence, galaxy density, deflection angle, respectively), and this mode coupling gives rise to small amounts of shear B-modes. As the resulting B-mode power spectra are negligible even for the next generation of weak lensing surveys, we focus here on the (larger) leading E-mode corrections. 
 \paragraph*{Reduced Shear}
 As the intrinsic size of source galaxies is unknown, galaxy-shape distortions measure the reduced shear, \mbox{$g =\gamma/(1+\kappa)$} rather than $\gamma$. The leading correction is
 \be
 g_{\alpha,\mathrm{obs}}^i(\hat{\mathbf n}) = \frac{\gamma_{\alpha, \mathrm{obs}}^i(\hat{\mathbf n})}{1-\kappa^i(\hat{\mathbf n})}\approx \gamma_{\alpha,\mathrm{obs}}(\hat{\mathbf n}) + \gamma_{\alpha,\mathrm{obs}}(\hat{\mathbf n})\kappa^i(\hat{\mathbf n})\,.
 \ee
The corresponding correction to the E-mode power spectrum is \citep{Dodelson06,Shapiro09}
\begin{align}
\nonumber \Delta C_{{EE}}^{ij}(\ell) =& \int \frac{d\chi}{\chi^4}\,W^i_{\kappa,\mathrm{s}}(\chi)W^j_{\kappa,\mathrm{s}}(\chi)\left[W^i_{\kappa,\mathrm{s}}(\chi)+W^j_{\kappa,\mathrm{s}}(\chi)\right]\\
\nonumber &\times \int\frac{d^2 L}{(2\pi)^2} \cos(2\phi_{L})
B_\mathrm{m}\left(\frac{\mathbf{L}}{\chi},\frac{\boldsymbol{\ell}-\mathbf{L}}{\chi},\frac{-\boldsymbol{\ell}}{\chi},\chi\right)\\
&\nonumber +\int\frac{d^2 L}{(2\pi)^2}\cos(2\phi_{L})\left[\cos(2\phi_{L})
+\cos(2\phi_{\mathbf{L}-\boldsymbol{\ell}})
\right]\\
\nonumber&\quad\times\,
C_{\kappa\kappa}^{ij}(L)\,C_{\kappa\kappa}^{ij}(|\mathbf{L}-\boldsymbol{\ell}|)\\
&+\left[C_{\kappa\kappa}^{ii}(\ell)\int\frac{dL}{2\pi}L\, C_{\kappa\kappa}^{jj}(L) +\left(i \leftrightarrow j\right)\right]
\,, 
\label{eq:dCEE1}
\end{align}
with $B_\mathrm{m}(\mathbf{k_1},\mathbf{k_2},\mathbf{k_3},\chi)$ the matter bispectrum at redshift $z(\chi)$, and where we have omitted the reduced shear effect on $\gamma_{\mathrm{IA}}$. While the three terms in Eq.~(\ref{eq:dCEE1}) are of the same order in the linear matter density contrast, the last two terms containing products of two angular convergence power spectra are strongly suppressed as they include an additional lens efficiency factor.
The B-mode power spectrum from reduced shear is given by convolution of two angular convergence power spectra with different phase factor (there is no bispectrum term as the first-order B-mode shear is zero); as it is known to be insignificant \citep{lensingbias,KH10}, we will not consider it further here.
For galaxy-galaxy lensing,
\begin{align}
\nonumber \Delta C^{ij}_{\delta_{\mathrm{l}}E}(\ell) =& \int \frac{d\chi}{\chi^4}\left[b^i_{1,\mathrm{l}}W^i_{\delta,\mathrm{l}}(\chi) + C^i_\mathrm{l}W^i_{\kappa,\mathrm{l}}(\chi) \right]\left(W^j_{\kappa,\mathrm{s}}(\chi)\right)^2\\
&\times \int\frac{d^2 L}{(2\pi)^2} \cos(2\phi_{L})
B_\mathrm{m}\left(\frac{\mathbf{L}}{\chi},\frac{\boldsymbol{\ell}-\mathbf{L}}{\chi},\frac{-\boldsymbol{\ell}}{\chi},\chi\right)\,.
\end{align}
 \paragraph*{Source Clustering and Magnification}
 Since the shear field is only observed at the positions of source galaxies, we now account for deviations from the mean source redshift distribution along specific lines of sight due to the physical (3D) clustering of sources and modulation of the selection function by magnification:
 \begin{align}
    n_{\mathrm{s}}^i(\chi)\rightarrow n_{\mathrm{s}}^i(\chi)
\left[1+{\delta^{(3\mathrm D)}_{\mathrm{s}}(\hat{\mathbf n} \chi, \chi)}
\textcolor{black}{} \right]\left[1+ C_\mathrm{s}^i \kappa\left(\hat{\mathbf n}, z \right)\right]\,.
\label{eq:deltan}
 \end{align}
Here we have introduced the convergence field of a source plane at redshift $z$
 \begin{align}
\nonumber \kappa\left(\hat{\mathbf n}, z \right) &= \frac{3 \om H_0^2}{2c^2}\int_0^{z(\chi)} d\chi \,
\frac{\chi}{a(\chi)} \frac{\chi(z)-\chi}{\chi(z)} \delta_\mathrm{m}(\hat{\mathbf n}\chi,\chi)\\
&= \int_0^{\chi(z)} d\chi \,W_\kappa(\chi,z) \,\delta_\mathrm{m}(\hat{\mathbf n}\chi,\chi)\,,
 \end{align}
 where in the second step we defined $W_\kappa(\chi,z)$, the lens efficiency of a source plane at redshift $z$. 
  
The corresponding corrections to the shear field are computed starting from the deflection tensor (Eq.~\ref{eq:Psi}) with the modified redshift distribution,
\begin{eqnarray}
\nonumber \Delta \Psi_{\alpha\beta}(\bs{\ell})&=&2\int\!\! \frac{d^2L}{(2\pi)^2} \int\! d\chi\; n_{\mathrm{s}}^i(\chi)\Bigg[ b_{1,\mathrm{s}}^i  \delta_{\mathrm{m}}((\bs{\ell}- \mathbf{L})/\chi, \chi)\\
\nonumber&&+C_\mathrm{s}^i\int_0^\chi\!\!\! d\chi''W_\kappa(\chi'',z(\chi))\delta_{\mathrm{m}}((\bs{\ell}- \mathbf{L})/\chi'', \chi'') \Bigg]\\
 && \times \int_0^\chi\!\!\!  d \chi' W_\kappa(\chi',z(\chi)) \frac{L_\alpha L_\beta}{L^2} \delta_\mathrm{m}\left(\mathbf{L}/\chi', \chi'\right),
\end{eqnarray} 
where we omitted the $\mathcal{O}\left(\delta_{\mathrm{m}}^3\right)$ term as the corresponding $\mathcal{O}\left(\delta_{\mathrm{m}}^4\right)$ power spectrum correction vanishes after carrying out the $\phi_L$ integral. Projection into shear components (Eqs.~\ref{eq:shear_components}+\ref{eq:EB}) yields
\begin{eqnarray}
\nonumber \Delta\gamma_ E^i (\bs{\ell})&=& 
\!\!\int\!\! \frac{d^2L}{(2\pi)^2} \int\! d\chi\; {n}_{\mathrm{s}}^i(\chi)\Bigg[ b_{1,\mathrm{s}}^i  \delta_{\mathrm{m}}((\bs{\ell}- \mathbf{L})/\chi, \chi)\\
\nonumber  &&  +C_\mathrm{s}^i\int_0^\chi\!\!\! d\chi''W_\kappa(\chi'',z(\chi))\delta_{\mathrm{m}}((\bs{\ell}- \mathbf{L})/\chi'', \chi'') \Bigg]\\
\nonumber &&  \times \underbrace{T_\alpha(\bs{\ell}) T_\alpha(\mathbf L)}_{\cos(2 \phi_L)}  \int_0^\chi\!\!\! d \chi' W_\kappa(\chi',z(\chi))
\delta_\mathrm{m}\left(\mathbf{L}/\chi', \chi'\right).\\
\label{eq:g_E}
\end{eqnarray}
The correction $ \Delta\gamma_B$ is of the same form as Eq.~\ref{eq:g_E}, with $T_\alpha(\mathbf l) T_\alpha(\mathbf L)\rightarrow \epsilon_{\alpha\beta}T_\alpha(\mathbf l) T_\beta(\mathbf L) =\sin(2 \phi_L)$.

The resulting power spectrum corrections are then straightforward to calculate. While the source clustering fluctuations in Eq.~\ref{eq:deltan} are orders of magnitudes larger than the magnification fluctuations, several of the corresponding power spectrum correction terms are suppressed by the lens efficiency being zero at the source plane ($W_\kappa(\chi(z),z) =0$). There are three different (22)-type corrections to $C_{{EE/BB}}$ permitted under the Limber approximation. As for reduced shear, these terms are much smaller than the bispectrum terms, and we show here only the dominant term, proportional to the source clustering power spectrum, 
\begin{widetext}
\begin{align}
\nonumber \Delta C^{ij}_{{EE}}(\ell) = 
 &\;C_{\mathrm{s}}^i \int  d\chi \, n_{\mathrm{s}}^i(\chi) \int_0^\chi\! \frac{d\chi'}{\chi'^4}W_{\kappa,\mathrm{s}}^j(\chi')
W^2_\kappa\left(\chi',z(\chi)\right) \int\!\frac{d^2 L}{(2\pi)^2} \cos(2\phi_L) 
B_\mathrm{m}\!\left(\frac{\mathbf{L}}{\chi'},\frac{\bs{\ell}-\mathbf{L}}{\chi'},\frac{-\bs{\ell}}{\chi'},\chi'\right) + \left(i \leftrightarrow j\right)\\
\label{eq:dCEE} &+\int d\chi\! \frac{b_{1,\mathrm{s}}^i b_{1,\mathrm{s}}^j n_{\mathrm{s}}^i(\chi)n_{\mathrm{s}}^j(\chi)}{\chi^2}  \int\!\frac{d^2 L}{(2\pi)^2} \cos^2(2\phi_L) P_{\mathrm{m}}\left(\frac{|\bs{\ell}- \mathbf{L}|}{\chi}, \chi\right)\int_0^\chi d\chi'
\frac{W^2_\kappa\left(\chi',z(\chi)\right)}{\chi'^2} P_{\mathrm{m}}\left(\frac{L}{\chi'}, \chi'\right)+\cdots\\
\Delta C^{ij}_{{BB}}(\ell) = &\int\! d\chi \frac{b_{1,\mathrm{s}}^i b_{1,\mathrm{s}}^j n_{\mathrm{s}}^i(\chi)n_{\mathrm{s}}^j(\chi)}{\chi^2}  \int\!\frac{d^2 L}{(2\pi)^2} \sin^2(2\phi_L) P_{\mathrm{m}}\left(\frac{|\bs{\ell}- \mathbf{L}|}{\chi}, \chi\right)\int_0^\chi\! d\chi'
\frac{W^2_\kappa\left(\chi',z(\chi)\right)}{\chi'^2} P_{\mathrm{m}}\left(\frac{L}{\chi'}, \chi'\right)+\cdots \label{eq:dCBB}\\
\Delta C^{ij}_{\delta_{\mathrm{l}}E}(\ell) =& C_{\mathrm{s}}^j \int\! d\chi\, n_{\mathrm{s}}^j(\chi)\int_0^\chi\!  \frac{d\chi'}{\chi'^4}\left[b^i_{1,\mathrm{l}}W^i_{\delta,\mathrm{l}}(\chi') + C^i_\mathrm{l}W^i_{\kappa,\mathrm{l}}(\chi') \right] W^2_\kappa\left(\chi',z(\chi)\right) 
\int\!\frac{d^2 L}{(2\pi)^2} \cos(2\phi_{L})
B_\mathrm{m}\!\left(\frac{\mathbf{L}}{\chi'},\frac{\mathbf{l}-\mathbf{L}}{\chi'},\frac{-\mathbf{l}}{\chi'},\chi'\right).\label{eq:dCggl}
\end{align}
\end{widetext}
One can read off two useful observations about source clustering from these expressions: first, under the Limber approximation there notably are no bispectrum-type corrections from physical source (or lens-source) clustering, which would originate from the $\delta_\mathrm{s}^{(3\mathrm D)}$ term in Eq.~\ref{eq:deltan}, but are suppressed by the source plane lens efficiency. This is in contrast to the projected source clustering ansatz $\gamma^i(\hat{\mathbf n}) \rightarrow \gamma^i(\hat{\mathbf n})\left(1+\delta^i_{\mathrm{s,obs}}(\hat{\mathbf n}\right)$ \citep{lensingbias}. For broad tomographic source redshift bins, as applicable to DES-Y3, their approach gives rise to a spurious $\gamma^i\delta^i_{\mathrm{s,D}}$ term, which would cause substantial parameter biases for the DES-Y3 analysis. In the limit of narrow tomographic bins \citep[considered by][]{lensingbias}, both approaches yield the same power spectrum corrections.
Second, $\mathcal{O}\left(\delta_\mathrm{m}^4\right)$ lens-source clustering corrections to the galaxy-galaxy lensing power spectrum (Eq.~\ref{eq:dCggl}) vanish under the Limber approximation, which is relevant to the discussion of correlation function estimator effects\footnote{Going beyond linear galaxy clustering, there are $\mathcal{O}\left(\delta_\mathrm{m}^4\right)$ corrections to galaxy-galaxy lensing; these are contractions of nonlinear lens biasing with source magnification corrections, not 3D lens-source clustering.}.

The DES-Y3 analysis adopts pair-based estimators \citep{LS93,Schneider2002,Sheldon2004} for the 2PCF measurements, with weighting schemes \citep{y3-galaxyclustering,y3-gglensing,y3-cosmicshear1} (for clustering, galaxy-galaxy lensing, and cosmic shear, respectively). These estimators normalize the galaxy-galaxy lensing/cosmic shear 2PCF measurements by the observed (weighted) lens-source/source-source pair counts in each angular bin, which introduces a modulation by the \emph{projected} lens-source/source-source clustering correlation functions. For galaxy-galaxy lensing, this modulation describes the dilution by excess (clustered) lens-source pairs at the lens redshift; the DES Y3 galaxy--galaxy lensing analysis \citep{y3-gglensing} corrects for it at the measurement level through so-called Boost factors $B^{ij}(\theta)\approx 1 +w_{\rm{ls}}^{ij}(\theta)$ \citep{Hirata04,Mandelbaum05}. This effect is separate from the source clustering effect derived here, as is evident from the absence of $\mathcal{O}\left(\delta_\mathrm{m}^4\right)$ terms in Eq.~(\ref{eq:dCggl}).

The measured cosmic shear correlation functions $\xi_\pm$ are not debiased for the modulation of the observed pair counts by $(1+w_{\rm{ss}}^{ij}(\theta))$ from the pair-based estimator \citep{Peter_Bmodes,lensingbias}. This effect is partially cancelled by the (22)-type corrections to $C_{{EE/BB}}$ shown in Eqs.~(\ref{eq:dCEE},\ref{eq:dCBB}): for $\xi+$, in the limit of narrow source redshift bins (and hence restricting to auto-tomography combinations, $i=j$), the sum of these two corrections simplifies to the convolution of the convergence and angular source clustering power spectra
\begin{equation}
\Delta C^{(\mathcal{O}(\delta_{\mathrm{m}}^4))}_{{EE}}(\ell)+\Delta C_{{BB}}(\ell) 
\approx \left[C_{\kappa\kappa}* C_{\delta_{\rm{s}}\delta_{\rm{s}}}\right](\ell)\,,
\end{equation}
corresponding to $\Delta \xi_+(\theta) =  \xi_+(\theta)w_{\mathrm{ss}}(\theta)$ in configuration space \footnote{This expression is valid on small angular scales, where the transformations for density correlations (Eq.~\ref{eq:transform_w}) and $\xi_+$ (Eq.~\ref{eq:transform_xi}) have the same limit.}. 

\begin{figure*}
    \centering
    \includegraphics[width =0.98\textwidth]{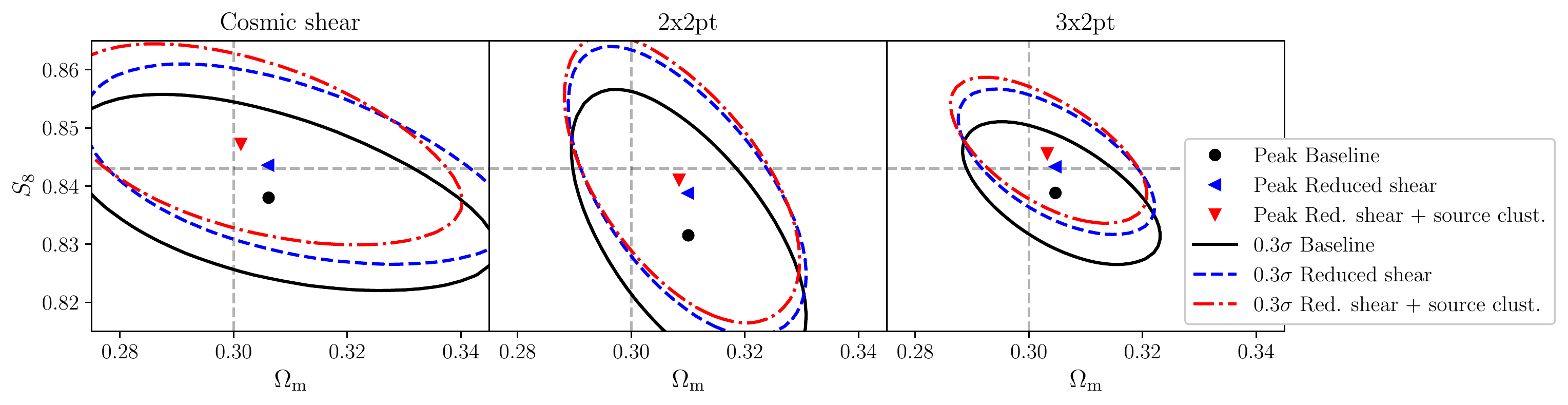}
    \caption{Robustness of parameter constraints to higher-order lensing effects. The offset between the blue/red contour and black contour illustrates the systematic bias incurred by neglecting reduced shear/reduced shear and source clustering terms. The contours levels are $0.3\sigma_{2\rm{D}}$, symbols indicate the 2D (marginalized) peak.}
    \label{fig:test_nlo}
\end{figure*}

 We evaluate the combined effect of reduced shear and source clustering assuming $b^i_{1,\mathrm{s}}=1$ and source magnification coefficients $C_\mathrm{s}^{1\cdots4}=\left(-1.17,-0.64,-0.55,0.80\right)$, estimated from data \citep{y3-2x2ptmagnification}, and modeling the matter bispectrum as tree-level but replacing the linear matter power with the nonlinear matter power spectrum \citep[see][for a detailed analysis of bispectum models]{Lazanu15}. The combined impact of higher-order lensing effects on parameter estimates is shown in Fig.~\ref{fig:test_nlo} for DES-Y3 $\Lambda$CDM analyses; the two-dimensional parameter bias is below $0.15\sigma_{2\mathrm{D}}$ for all analyses. 
 
We note that impact of reduced shear and source clustering on cosmic shear and galaxy-galaxy lensing is sensitive to squeezed-limit bispectrum configurations and thus the nonlinear bispectrum model. Accurate modeling of these effects will be required for weak lensing analyses of future surveys.
 
\section{Summary and Conclusions}
\label{sec:conc}
In order to obtain the best possible cosmological parameter constraints, the accuracy of analysis methodology has to be matched to the precision enabled by the data set. Hence the required accuracy depends on the survey properties, the combination of cosmological probes, and the model/parameter space under investigation. The analysis choices that must be made as part of the methodology development range from modeling that is critical to capturing the underlying physics, such as intrinsic alignments, to extensive validation of scale cuts imposed on the data vector, to technical decisions such as prior ranges or the precision of integration and interpolation routines that are employed in the modeling. The limiting factor in some cases is computing time; while of course the pipeline must be sufficiently accurate to match the precision of the data, it is also clear that increasing the complexity of the pipeline at a disproportionately high cost in computing time is unwise if the improvements are well below the statistical errors.  

In this paper we describe the cosmology and astrophysics modeling strategy and validation for the DES Y3 joint weak lensing and galaxy clustering analysis. We demonstrate that our analysis choices are sufficient to mitigate systematic biases in the DES Y3 parameter constraints due to theoretical modeling simplifications and uncertainties.

Our modeling systematics mitigation strategy can be summarized as a combination of (1) modeling and marginalizing over systematic-error parameterizations and (2) imposing scale cuts that exclude data points where unmodeled systematics are prevalent. In terms of modeling and marginalizing, we choose a baseline analysis parameterization that models the nonlinear matter power spectrum, galaxy bias, intrinsic alignments, lens magnification, and non-local contributions to galaxy--galaxy lensing. 
While necessarily incomplete, this model is sufficiently complex and robust for the DES Y3 3$\times$2pt analysis.
Specifically, we demonstrate through a large number of simulated likelihood analyses that (1) angular scale cuts mitigate the leading known, but unmodeled, systematic effects, and (2) the baseline parameterization is sufficiently flexible to marginalize over plausible variations from these parameterization choices. We further validate our numerical implementation by comparing independently-developed modeling pipelines, and find them to be in excellent mutual agreement. We conclude that the cosmology and astrophysics modeling is robust within the analysis choices derived here, and that theoretical modeling systematics are insignificant for DES Y3 cosmic shear, 2$\times$2pt and 3$\times$2pt analyses, as well as their combinations with external data. To our knowledge, this work presents the most detailed validation of a large-scale structure survey to date.

The dominant systematics limiting the constraining power of the DES Y3 baseline analyses are astrophysical effects in the nonlinear regime, in particular nonlinear galaxy bias, baryonic effects on the matter power spectrum, and non-local contributions from the one-halo and one-to-two halo transition regime to the large-scale galaxy--galaxy lensing signal. The former two set the angular scale cuts for the baseline analysis, excluding a substantial fraction of the measured signal-to-noise in the two-point correlation functions, while the latter degrades the constraining power of small-scale galaxy--galaxy lensing. 

The modeling challenges described in this paper have long been recognized. Successfully resolving them will allow us to achieve the ultimate goal of optimally extracting information from galaxy surveys, particularly from small spatial scales where a large but challenging-to-model signal resides. While our work here represents important developments in mitigating these uncertainties, substantial progress remains possible with future effort. For instance, suitable modeling or mitigating of baryonic effects on small scales can be used to extract more information from weak lensing surveys, as demonstrated by \citet{Asgari_2020} and \citet{hem2020} with recent data sets. Another example is nonlinear galaxy bias: while its modeling has become a standard tool in analyses of galaxy clustering in redshift space \citep[e.g.,][]{Sanchez2016,DAmico2020,Ivanov2020}, the application to angular clustering measurements remains challenging due to parameter degeneracies and projection effects. Progress in that particular direction for the DES Y3 analysis is being reported by \citet{y3-2x2ptbiasmodelling} and \citet{y3-simvalidation}, who validate an extension of the baseline analysis described here using next-to-leading-order galaxy biasing; the corresponding gains in constraining power are presented in Ref.~\citep{y3-3x2ptkp}. Due to additional nuisance parameters, the gain in constraining power from nonlinear bias modeling is significantly smaller than a mode counting argument might suggest.

Upcoming, more constraining, analyses will continue to balance model accuracy (and hence model complexity) with increased statistical constraining power. One of the main limiting factors for such
ambitious analyses will be the availability of sufficient validation tests, such as high-resolution, high-volume mock catalogs that include plausible variations of the relevant nonlinear astrophysics, such as baryonic feedback, or variations in the halo--galaxy connection. This issue will be front-and-center for future analyses of data from Rubin Observatory's LSST, the Euclid mission, and the Roman Space Telescope, in particular when combining the constraining power of multiple surveys' data.

Future analyses will need to evaluate the trade-offs that arise from increasing the number of nuisance parameters in the theoretical models.  Model complexification can improve the accuracy of cosmological inference by bringing the model space closer to the true Universe.  But this may come at the price of increased parameter degeneracies and reduced cosmological precision, inference instabilities due to multiple minima,\footnote{See Appendix B of \citep{y3-cosmicshear1} for a characterization of bimodality in the TATT parameter space due to noise.} and biased confidence contours due to prior-volume projections. 
It is therefore important to develop the \textit{least} complex analysis that is robust for any given science case and data set, and to attend closely to priors on additional parameters.  Mapping out the modeling trade space will require a large set of simulated analyses customized to the noise level, probe combination, and cosmological model space of a given experiment.  Advances in theory that restrict rather than expand the model space will be extremely valuable to to achieve high-accuracy cosmology.

\section*{Acknowledgements}
EK thanks Fabian Schmidt and Masahiro Takada for helpful discussions, and Fabian Schmidt for feedback on an early draft. EK is supported in part by the Department of Energy grant DE-SC0020247, the David \& Lucile Packard Foundation and the Alfred P. Sloan Foundation. XF is supported by the Department of Energy grant DE-SC0020215.

This work made use of the software packages {\tt PolyChord} \citep{polychord1,polychord2}, {\tt GetDist} \citep{getdist}, {\tt matplotlib} \citep{matplotlib}, and {\tt numpy} \citep{numpy}.

\input{des_acknowledgements}
\bibliography{y3kp.bib,methods.bib}

\label{lastpage}
\end{document}

%% file: authorlist.tex

\author{E.~Krause}\email{krausee@arizona.edu}
\affiliation{Department of Astronomy/Steward Observatory, University of Arizona, 933 North Cherry Avenue, Tucson, AZ 85721-0065, USA} 
\author{X.~Fang}
\affiliation{Department of Astronomy/Steward Observatory, University of Arizona, 933 North Cherry Avenue, Tucson, AZ 85721-0065, USA}
\author{S.~Pandey}
\affiliation{Department of Physics and Astronomy, University of Pennsylvania, Philadelphia, PA 19104, USA}
\author{L.~F.~Secco}
\affiliation{Department of Physics and Astronomy, University of Pennsylvania, Philadelphia, PA 19104, USA}
\affiliation{Kavli Institute for Cosmological Physics, University of Chicago, Chicago, IL 60637, USA}
\author{O.~Alves}
\affiliation{Department of Physics, University of Michigan, Ann Arbor, MI 48109, USA}
\affiliation{Instituto de F\'{i}sica Te\'orica, Universidade Estadual Paulista, S\~ao Paulo, Brazil}
\affiliation{Laborat\'orio Interinstitucional de e-Astronomia - LIneA, Rua Gal. Jos\'e Cristino 77, Rio de Janeiro, RJ - 20921-400, Brazil}
\author{H.~Huang}
\affiliation{Department of Physics, University of Arizona, Tucson, AZ 85721, USA}
\author{J.~Blazek}
\affiliation{Department of Physics, Northeastern University, Boston, MA 02115, USA}
\affiliation{Laboratory of Astrophysics, \'Ecole Polytechnique F\'ed\'erale de Lausanne (EPFL), Observatoire de Sauverny, 1290 Versoix, Switzerland}
\author{J.~Prat}
\affiliation{Department of Astronomy and Astrophysics, University of Chicago, Chicago, IL 60637, USA}
\affiliation{Kavli Institute for Cosmological Physics, University of Chicago, Chicago, IL 60637, USA}
\author{J.~Zuntz}
\affiliation{Institute for Astronomy, University of Edinburgh, Edinburgh EH9 3HJ, UK}
\author{T.~F.~Eifler}
\affiliation{Department of Astronomy/Steward Observatory, University of Arizona, 933 North Cherry Avenue, Tucson, AZ 85721-0065, USA}
% ...

\collaboration{DES Collaboration}

%% file: des_acknowledgements.tex
Funding for the DES Projects has been provided by the U.S. Department of Energy, the U.S. National Science Foundation, the Ministry of Science and Education of Spain, 
the Science and Technology Facilities Council of the United Kingdom, the Higher Education Funding Council for England, the National Center for Supercomputing 
Applications at the University of Illinois at Urbana-Champaign, the Kavli Institute of Cosmological Physics at the University of Chicago, 
the Center for Cosmology and Astro-Particle Physics at the Ohio State University,
the Mitchell Institute for Fundamental Physics and Astronomy at Texas A\&M University, Financiadora de Estudos e Projetos, 
Funda{\c c}{\~a}o Carlos Chagas Filho de Amparo {\`a} Pesquisa do Estado do Rio de Janeiro, Conselho Nacional de Desenvolvimento Cient{\'i}fico e Tecnol{\'o}gico and 
the Minist{\'e}rio da Ci{\^e}ncia, Tecnologia e Inova{\c c}{\~a}o, the Deutsche Forschungsgemeinschaft and the Collaborating Institutions in the Dark Energy Survey. 

The Collaborating Institutions are Argonne National Laboratory, the University of California at Santa Cruz, the University of Cambridge, Centro de Investigaciones Energ{\'e}ticas, 
Medioambientales y Tecnol{\'o}gicas-Madrid, the University of Chicago, University College London, the DES-Brazil Consortium, the University of Edinburgh, 
the Eidgen{\"o}ssische Technische Hochschule (ETH) Z{\"u}rich, 
Fermi National Accelerator Laboratory, the University of Illinois at Urbana-Champaign, the Institut de Ci{\`e}ncies de l'Espai (IEEC/CSIC), 
the Institut de F{\'i}sica d'Altes Energies, Lawrence Berkeley National Laboratory, the Ludwig-Maximilians Universit{\"a}t M{\"u}nchen and the associated Excellence Cluster Universe, 
the University of Michigan, the National Optical Astronomy Observatory, the University of Nottingham, The Ohio State University, the University of Pennsylvania, the University of Portsmouth, 
SLAC National Accelerator Laboratory, Stanford University, the University of Sussex, Texas A\&M University, and the OzDES Membership Consortium.

The DES data management system is supported by the National Science Foundation under Grant Numbers AST-1138766 and AST-1536171.
The DES participants from Spanish institutions are partially supported by MINECO under grants AYA2015-71825, ESP2015-88861, FPA2015-68048, SEV-2012-0234, SEV-2016-0597, and MDM-2015-0509, 
some of which include ERDF funds from the European Union. IFAE is partially funded by the CERCA program of the Generalitat de Catalunya.
Research leading to these results has received funding from the European Research
Council under the European Union's Seventh Framework Program (FP7/2007-2013) including ERC grant agreements 240672, 291329, and 306478.
We  acknowledge support from the Australian Research Council Centre of Excellence for All-sky Astrophysics (CAASTRO), through project number CE110001020.

This manuscript has been authored by Fermi Research Alliance, LLC under Contract No. DE-AC02-07CH11359 with the U.S. Department of Energy, Office of Science, Office of High Energy Physics. The United States Government retains and the publisher, by accepting the article for publication, acknowledges that the United States Government retains a non-exclusive, paid-up, irrevocable, world-wide license to publish or reproduce the published form of this manuscript, or allow others to do so, for United States Government purposes.

Based in part on observations at Cerro Tololo Inter-American Observatory, 
National Optical Astronomy Observatory, which is operated by the Association of 
Universities for Research in Astronomy (AURA) under a cooperative agreement with the National 
Science Foundation.